\documentclass[a4paper,11pt]{article}
\pdfoutput=1

\usepackage{jheppub}

\usepackage{graphicx}
\usepackage{hyperref}
\usepackage{orcidlink}

\def\bd{B^0}
\def\bs{B_s^0}

\def \beq{\begin{equation}}
\def \eeq{\end{equation}}
\def \bea{\begin{eqnarray}}
\def \eea{\end{eqnarray}}

\def \s{\sqrt{2}}
\def \({\left(}
\def \){\right)}
\def \[{\left[}
\def \]{\right]}
\def \nn{\nonumber}

\def \de{\delta}
\def \la{\lambda}

\def\btopik{B\to\pi K}

\allowdisplaybreaks

\title{\boldmath Isospin-based EWP-tree Relations}

\author[a]{Bhubanjyoti Bhattacharya\,\orcidlink{0000-0003-2238-321X},}
\author[b]{Marianne Bouchard\,\orcidlink{0009-0005-2885-7473},}
\author[b,c]{Alexandre Jean\,\orcidlink{0009-0005-2354-3299},}
\author[b]{\\ David London\,\orcidlink{0000-0002-4407-5624},}
\author[b]{and Ipsita Ray\,\orcidlink{0000-0002-5506-754X}}

\affiliation[a]{Department of Natural Sciences, Lawrence Technological University, Southfield, MI 48075, USA}
\affiliation[b]{Physique des Particules, Universit\'e de Montr\'eal, 1375 Avenue Th\'er\`ese-Lavoie-Roux, Montr\'eal, QC, Canada  H2V 0B3}
\affiliation[c]{Institute for Particle Physics Phenomenology, Department of Physics, Durham University, Durham DH1 3LE, U.K.}

\emailAdd{bbhattach@ltu.edu}
\emailAdd{marianne.bouchard.5@umontreal.ca}
\emailAdd{alexandre.jean.1@umontreal.ca}
\emailAdd{london@lps.umontreal.ca}
\emailAdd{ipsita.ray@umontreal.ca}

\date{\today}

\preprint{UdeM-GPP-TH-25-307, IPPP/25/67}

\abstract{
In 1998, it was shown that, if flavor SU(3) symmetry [SU(3)$_F$] is assumed in charmless $B \to PP$ decays ($P$ is a light pseudoscalar meson), some reduced matrix elements involving electroweak penguin (EWP) operators are related to those involving tree operators. Similarly, EWP diagrams are related to tree diagrams. These SU(3)$_F$ EWP-tree relations were recently used in global analyses of $B \to PP$ decays. They have also been used over the years in analyses of the $\btopik$ puzzle, even though the $\btopik$ amplitudes are related by isospin symmetry [SU(2)$_I$], and not the full SU(3)$_F$. In this paper, we show that, even if only SU(2)$_I$ is assumed, there are still EWP-tree relations. In $\Delta S=0$ decays, these relations are similar to those of SU(3)$_F$, and can be used to take into account the EWP contributions in the extraction of the CP phase $\alpha$ from $B \to \pi\pi$ decays. In $\Delta S=1$ decays, the SU(2)$_I$ EWP-tree relations are quite different from those of SU(3)$_F$; when these are used to analyze the $\btopik$ puzzle, one now finds a 4-5$\sigma$ discrepancy with the Standard Model, much larger than what was previously found. We argue that, if one analyzes a set of hadronic $B$ decays whose amplitudes are related by isospin, one must use the SU(2)$_I$ EWP-tree relations for that set of decays in the analysis.}

\begin{document}

\maketitle
\flushbottom

\section{Introduction}

Over the years, there has been a great deal of interest in two-body hadronic $B$ decays. At first, this was because CP violation could be measured in such decays. Later, as more and more measurements were made, it was to try to analyze the data, to see if they were consistent with the Standard Model (SM).

Although great progress has been made in calculating hadronic $B$ decays, there is still a good deal of uncertainty in the results of such computations. For this reason, certain tools have been used to relate different processes, permitting the joint analysis of the various observables in these decays without theoretical input. One such tool is group theory. For example, the amplitudes for the three decays $B^+ \to \pi^+ \pi^0$, $\bd\to\pi^+\pi^-$ and $\bd\to\pi^0\pi^0$ obey a triangle relation under isospin [SU(2)$_I$]. The analysis of the branching ratios and CP asymmetries in these decays allows one to extract the CP phase $\alpha$
\cite{Gronau:1990ka}. More generally, all charmless $B \to PP$ decays, where $P$ is a light pseudoscalar meson, are related to one another by flavor SU(3) symmetry [SU(3)$_F$]. That is, using the SU(3)$_F$ Wigner-Eckart theorem, all $B \to PP$ amplitudes can be expressed in terms of the same SU(3)$_F$ reduced matrix elements (RMEs). This allows one to perform fits to all the $B \to PP$ data, to test their internal consistency and to see how they compare with the predictions of the SM. For recent analyses that use RMEs, see Refs.~\cite{Berthiaume:2023kmp,Bhattacharya:2025wcq}. Older analyses that do not use RMEs and often rely on theoretical input can be found in Refs.~\cite{Chiang:2004, Fleischer:2007,Cheng:2014}.

The weak Hamiltonian that governs charmless $B \to PP$ decays is \cite{Buchalla:1995vs}
\beq
H_W = \frac{G_F}{\sqrt{2}} \sum_{q=d,s}\left(\lambda_u^{(q)} \sum_{i=1}^{2} c_i Q_i^{(q)}
- \lambda_t^{(q)} \sum_{i=3}^{10} c_i Q_i^{(q)} \right) ~.
\label{eq:HW}
\eeq
The $c_i$s ($i = 1$-10) are Wilson coefficients. $\lambda_{u}^{(q)} \equiv V_{ub}^* V_{uq}$ and $\lambda_{t}^{(q)} \equiv V_{tb}^* V_{tq}$ ($q=d,s$), where the $V_{ij}$ are elements of the Cabibbo-Kobayashi-Maskawa matrix. The operators proportional to $\lambda_u^{(q)}$ are tree operators:
\beq
Q_1^{(q)} = ({\bar b}u)_{V-A}({\bar u}q)_{V-A} ~~,~~~~
Q_2^{(q)} = ({\bar b}q)_{V-A}({\bar u}u)_{V-A} ~.
\label{eq:Q1,2operators}
\eeq
The operators proportional to $\lambda_t^{(q)}$ include both gluonic penguin and electroweak penguin operators. The four gluonic penguin operators are defined as follows :
\begin{equation}
\begin{split}
    Q_{3}^{(q)}&=({\bar b}q)_{V-A}(({\bar u}u)_{V-A}+({\bar d}d)_{V-A}+({\bar s}s)_{V-A}) ~, \\
    Q_{4}^{(q)}&=({\bar b}u)_{V-A}({\bar u}q)_{V-A}+({\bar b}d)_{V-A}({\bar d}q)_{V-A}+({\bar b}s)_{V-A}({\bar s}q)_{V-A} ~, \\
    Q_{5}^{(q)}&=({\bar b}q)_{V-A}(({\bar u}u)_{V+A}+({\bar d}d)_{V+A}+({\bar s}s)_{V+A}) ~, \\
    Q_{6}^{(q)}&=({\bar b}u)_{V-A}({\bar u}q)_{V+A}+({\bar b}d)_{V-A}({\bar d}q)_{V+A}+({\bar b}s)_{V-A}({\bar s}q)_{V+A} ~.
    \label{eq:gluon_oper}
\end{split}
\end{equation}
The four electroweak penguin (EWP) operators are
\begin{equation}
\begin{split}
    Q_{7}^{(q)}&=\frac{3}{2}\left[({\bar b}q)_{V-A}\left(\frac{2}{3}({\bar u}u)_{V+A}-\frac{1}{3}({\bar d}d)_{V+A}-\frac{1}{3}({\bar s}s)_{V+A}\right)\right] ~, \\
    Q_{8}^{(q)}&=\frac{3}{2}\left[\frac{2}{3}({\bar b}u)_{V-A}({\bar u}q)_{V+A}-\frac{1}{3}({\bar b}d)_{V-A}({\bar d}q)_{V+A}-\frac{1}{3}({\bar b}s)_{V-A}({\bar s}q)_{V+A}\right] ~, \\
    Q_{9}^{(q)}&=\frac{3}{2}\left[({\bar b}q)_{V-A}\left(\frac{2}{3}({\bar u}u)_{V-A}-\frac{1}{3}({\bar d}d)_{V-A}-\frac{1}{3}({\bar s}s)_{V-A}\right)\right] ~, \\
    Q_{10}^{(q)}&=\frac{3}{2}\left[\frac{2}{3}({\bar b}u)_{V-A}({\bar u}q)_{V-A}-\frac{1}{3}({\bar b}d)_{V-A}({\bar d}q)_{V-A}-\frac{1}{3}({\bar b}s)_{V-A}({\bar s}q)_{V-A}\right] ~.
\end{split}
\label{eq:EWP_oper}
\end{equation}

For the electroweak penguin operators, the values of the Wilson
coefficients $c_7$ and $c_8$ are much smaller than those of $c_9$ and
$c_{10}$ \cite{Buchalla:1995vs}, so that they can be neglected. The
EWP operators are then only $Q_{9}$ and $Q_{10}$, which are purely
$(V-A) \times (V-A)$. They can be written as\footnote{A similar decomposition of $Q_9$ and $Q_{10}$ in terms of tree and penguin operators can be found in Ref.~\cite{Paz:2002ev}.}
\beq
Q_{9} = \frac32 Q_2 - \frac12 Q_3 ~~,~~~~ Q_{10} = \frac32 Q_1 - \frac12 Q_4 ~.
\label{EWPtreeops}
\eeq

As noted above, all charmless $B \to PP$ amplitudes can be written in terms of the SU(3)$_F$ RMEs $\langle PP || H_W || B \rangle$. It is useful to split these RMEs into two pieces, one proportional to $\lambda_u^{(q)}$, the other proportional to $\lambda_t^{(q)}$, doubling the number of independent RMEs. The $\lambda_u^{(q)}$ RMEs get contributions from the tree operators [Eq.~(\ref{eq:Q1,2operators})], while the $\lambda_t^{(q)}$ RMEs get contributions from both the gluonic penguin [Eq.~(\ref{eq:gluon_oper})] and EWP operators [Eq.~(\ref{eq:EWP_oper})].

Using Eq.~(\ref{EWPtreeops}), it is possible to write any RME generated by an EWP operator as the sum of two RMEs, one for each of its contributions from the tree and penguin operators.
Suppose now that there are some $\lambda_t^{(q)}$ RMEs to which the gluonic penguin operators do not contribute; these RMEs are generated only by the EWP operators.
From Eq.~(\ref{EWPtreeops}), once the contribution from the penguin operator is absent, the $\lambda_t^{(q)}$ RME coming from the EWP operator will be directly proportional to the corresponding $\lambda_u^{(q)}$ RME.
This is known as an EWP-tree relation (at the level of RMEs), and it has the effect of reducing the number of independent RMEs.

Another tool that has been greatly used in analyzing hadronic $B$ decays is topological diagrams \cite{Cheng:1986}. In
Refs.~\cite{Gronau:1994rj, Gronau:1995hn}, it was shown that the amplitudes for charmless $B \to PP$ decays can be written in terms of ten diagrams: $T$ (color-allowed tree), $C$ (color-suppressed tree), $P_{tc}$ and $P_{uc}$ (gluonic penguin), $P_{EW}$ (color-allowed EWP), $P_{EW}^C$ (color-suppressed EWP), $E$ (exchange), $A$ (annihilation), and $PA_{tc}$ and $PA_{uc}$ (penguin annihilation).  $T$, $C$, $P_{uc}$, $E$, $A$, and $PA_{uc}$ are all proportional to $\lambda_{u}^{(q)}$, while $P_{tc}$, $P_{EW}$, $P_{EW}^C$, and $PA_{tc}$ are all proportional to $\lambda_{t}^{(q)}$. Diagrams are equivalent to SU(3)$_F$ RMEs \cite{Gronau:1994rj}. That is, each RME can be written as a linear combination of diagrams. However, here there is a caveat: there are more diagrams than RMEs, so the diagrams are not all independent -- only certain combinations of diagrams (those that correspond to RMEs) appear in the amplitudes. The fact that RMEs can be written in terms of diagrams means that the RME EWP-tree relations can be converted into diagrammatic EWP-tree relations, relating the EWP diagrams to the tree diagrams.

The point of the above discussion is the following. If there are RMEs in charmless hadronic $B \to PP$ decays to which the penguin operators do not contribute, then there will be EWP-tree relations; these can be written at the level of RMEs or diagrams. So the question is: are there such RMEs? The answer is yes: in Ref.~\cite{Gronau:1998fn}, Gronau, Pirjol and Yan (GPY) extended the work of Neubert and Rosner \cite{Neubert:1998pt, Neubert:1998jq}, and derived EWP-tree relations in charmless $B \to PP$ decays in the context of SU(3)$_F$ symmetry.  These EWP-tree relations were used in Refs.~\cite{Berthiaume:2023kmp,Bhattacharya:2025wcq}, which performed fits to all the $B \to PP$ data under the assumption of SU(3)$_F$. (A significant discrepancy with the SM was found.)

Following the appearance of Ref.~\cite{Gronau:1998fn}, these same EWP-tree relations were most often used in the analysis of $\btopik$ decays. First, there is the so-called ``$\btopik$ puzzle.'' The four decays $B^+ \to \pi^+ K^0$, $B^+ \to \pi^0 K^+$, $\bd\to \pi^- K^+$, and $\bd\to\pi^0 K^0$ are related by isospin: within this symmetry, their amplitudes obey a quadrilateral relation. However, the measurements of the observables in these decays are not completely consistent with one another \cite{Buras:2003yc, Buras:2003dj, Buras:2004ub}. In order to quantify this discrepancy, the amplitudes for these decays are written in terms of diagrams, and a fit is performed, using the magnitudes and phases of these diagrams as the free parameters. This is where the EWP-tree relations enter: they are used in the fit to express the EWP diagrams in terms of the tree diagrams. A number of $\btopik$ fits have been performed over the years, and they have typically found that the data disagree with the SM at the level of 2-3$\sigma$ \cite{Baek:2004rp, Baek:2007yy, Baek:2009pa, Beaudry:2017gtw, Bhattacharya:2021shk, Datta:2024zrl}. Second, a relation among $\btopik$ observables was shown to hold approximately using the SU(3)$_F$ EWP tree-relations ($+$ theory input) \cite{Gronau:2005kz}.

But this raises a question. The EWP-tree relations of GPY were derived assuming the full SU(3)$_F$ symmetry. This means that, for example, the diagrams in $\btopik$ decays are the same as those in all other charmless $B\to PP$ decays. But the symmetry that relates the $\btopik$ amplitudes to one another is just isospin. Under isospin, the diagrams in $\btopik$ decays are {\it unrelated} to those in other decays. Can we not derive EWP-tree relations, applicable only to $\btopik$ decays, assuming only SU(2)$_I$?

In this paper, we show that, if only SU(2)$_I$ is assumed, there are still EWP-tree relations. Whereas the SU(3)$_F$ EWP-tree relations were derived using all $B \to PP$ decays, the SU(2)$_I$ EWP-tree relations are derived for six different sets of decays whose amplitudes are related by isospin. There are three sets of $\Delta S = 0$ decays ($B \to\pi\pi$, $B_s^0 \to\pi{\bar K}$, and $B \to K{\bar K}$) and three sets of $\Delta S = 1$ decays ($\btopik$, $B_s^0 \to K{\bar K}$, and $B_s^0 \to \pi\pi$). In general, the SU(2)$_I$ EWP-tree relations are different for different sets of decays.
We find that the SU(2)$_I$ EWP-tree relations for the sets of $\Delta S = 0$ decays are similar to some of those of SU(3)$_F$. The $B\to\pi\pi$ EWP-tree relations can be used to take EWP contributions into account in the extraction of $\alpha$ from $B \to \pi\pi$ decays. On the other hand, for the sets of $\Delta S=1$ decays, the SU(2)$_I$ EWP-tree relations are not the same as the SU(3)$_F$ EWP-tree relations. Using the SU(2)$_I$  EWP-tree relations for $\btopik$, we show that the relation among $\btopik$ observables that was previously shown to hold approximately using the SU(3)$_F$ EWP-tree relations ($+$ theory input), actually holds exactly under SU(2)$_I$. We also find that, when the $\btopik$ SU(2)$_I$ EWP-tree relations are used in the fit to the $\btopik$ data, the discrepancy with the SM turns out to be at the level of 4-5$\sigma$, quite a bit larger than what was found previously. The takeaway here is the following: if one is analyzing a set of hadronic $B$ decays whose amplitudes are related by isospin, the SU(2)$_I$ EWP-tree relations for that set of decays must be used in the analysis.

We begin in Sec.~\ref{section:EWP-tree su3} by deriving the EWP-tree relations assuming SU(3)$_F$.  In Sec.~\ref{section:EWP-tree su2 deltaS0}, we derive the SU(2)$_I$ EWP-tree relations for the three sets of $\Delta S = 0$ decays. A similar analysis for the three sets of $\Delta S =1$ decays is presented in Sec.~\ref{section:EWP-tree su2 deltaS1}. In Sec.~\ref{Sec:alphaBpipi}, we show how the $B \to\pi\pi$ SU(2)$_I$ EWP-tree relation can be incorporated into the extraction of $\alpha$ from $B\to\pi\pi$ decays. In Sec.~\ref{section:btopik}, we provide an update to the $\btopik$ puzzle by redoing the fits to the $\btopik$ data with both the SU(3)$_F$ and $\btopik$ SU(2)$_I$ EWP-tree relations, and comparing the results. We conclude in Sec.~\ref{section:conclusion}.

\section{\boldmath EWP-tree relations in SU(3)$_F$}
\label{section:EWP-tree su3}

In this section, we rederive the EWP-tree relations in the context of SU(3)$_F$ symmetry. Our derivation uses a somewhat different method than that of GPY \cite{Gronau:1998fn}.

The amplitudes for a number of different $B \to PP$ decays, where $B=(B^+, B^0, \bs)$ and $P$ is a light pseudoscalar meson, are related by the flavor SU(3) symmetry. Under SU(3)$_F$, the quarks $(u,d,s)$ form a triplet $\mathbf{3}$ and the antiquarks $({\bar d}, -{\bar u}, {\bar s})$ form an antitriplet $\mathbf{3^*}$. The initial-state $B$ is also a $\mathbf{3}$, while the final-state $P$s are members of the meson octet $\mathbf{8}$. All members of the $\mathbf{8}$ are considered to be identical spin-0 bosons, so the final $PP$ state must be symmetrized, with $(\mathbf{8} \otimes \mathbf{8})_{\rm sym} = \mathbf{1} \oplus \mathbf{8} \oplus \mathbf{27}$. The decay involves the quark transitions ${\bar b} \to {\bar q} u {\bar u}$ and ${\bar b} \to ({\bar q} d {\bar d}+{\bar q} s {\bar s})$ ($q = d, s$), so the weak Hamiltonian transforms as a $\mathbf{3^*}$, $\mathbf{6}$ or $\mathbf{15^*}$. This leads to five RMEs:
\bea
& \langle {\bf 1} || H_W({\bf 3^*}) || {\bf 3} \rangle ~~,~~~~
\langle {\bf 8} || H_W({\bf 3^*}) || {\bf 3} \rangle ~, & \nn\\
& \langle {\bf 8} || H_W({\bf 6}) || {\bf 3} \rangle ~~,~~~~
\langle {\bf 8} || H_W({\bf 15^*}) || {\bf 3} \rangle ~~,~~~~
\langle {\bf 27} || H_W({\bf 15^*})  || {\bf 3} \rangle ~, &
\label{5RMEs}
\eea
where $H_W({\bf X})$ represents the piece of the weak Hamiltonian that transforms as the representation ${\bf X}$ of SU(3)$_F$.

The weak Hamiltonian of Eq.~(\ref{eq:HW}) contains tree, gluonic penguin, and electroweak penguin operators. In order to derive the EWP-tree relations, it is necessary to determine how these operators transform under SU(3)$_F$. We note that operators of the form $({\bar b} q_1)({\bar q}_2 q_3)$ transform under the assumed symmetry group as ${\bar q}_1 {\bar q}_3 q_2$ (in this order) \cite{Gronau:1998fn}. For the group SU(3)$_F$, the tree operators $Q_1$ and $Q_2$ [Eq.~(\ref{eq:Q1,2operators})] transform as
\beq
Q_1 =
\begin{cases}
\frac{1}{2} \cdot {\bf 6}_{I = \frac{1}{2}} - \frac{1}{\sqrt{3}}\cdot{\bf 15^*}_{I = \frac{3}{2}} - \frac{1}{2\sqrt{6}}\cdot{\bf 15^*}_{I = \frac{1}{2}} -\frac{1}{2} \cdot {\bf 3^*}^a_{I = \frac{1}{2}} + \frac{1}{2\sqrt{2}}\cdot{\bf 3^*}^s_{I = \frac{1}{2}} ~~, & q=d ~, \\
-\frac{1}{2} \cdot {\bf 6}_{I = 1} - \frac{1}{2} \cdot {\bf 15^*}_{I = 1} - \frac{1}{2\sqrt{2}}\cdot{\bf 15^*}_{I = 0} -\frac{1}{2} \cdot {\bf 3^*}^a_{I = 0}  + \frac{1}{2\sqrt{2}}\cdot{\bf 3^*}^s_{I = 0} ~~, & q=s ~,
\end{cases} \label{TopsSU3F}
\eeq
\beq
Q_2 =
\begin{cases}
- \frac{1}{2} \cdot {\bf 6}_{I = \frac{1}{2}} - \frac{1}{\sqrt{3}}\cdot{\bf 15^*}_{I = \frac{3}{2}} - \frac{1}{2\sqrt{6}}\cdot{\bf 15^*}_{I = \frac{1}{2}} + \frac{1}{2} \cdot {\bf 3^*}^a_{I = \frac{1}{2}} + \frac{1}{2\sqrt{2}}\cdot{\bf 3^*}^s_{I = \frac{1}{2}} ~~, & q=d ~, \\
\frac{1}{2}\cdot {\bf 6}_{I = 1} - \frac{1}{2} \cdot {\bf 15^*}_{I = 1} - \frac{1}{2\sqrt{2}}\cdot{\bf 15^*}_{I = 0} + \frac{1}{2}\cdot {\bf 3^*}^a_{I = 0} + \frac{1}{2\sqrt{2}}\cdot{\bf 3^*}^s_{I = 0} ~~, & q=s ~,
\end{cases} \nn
\eeq
where $\mathbf{3^{*a}}$ corresponds to the antisymmetric part of the triplet (it is multiplied by $c_2 - c_1$) and $\mathbf{3^{*s}}$ is the symmetric part of the triplet (it is multiplied by $c_1 + c_2$).
All gluonic penguin operators [Eq.~(\ref{eq:gluon_oper})] transform purely as a ${\bf 3^*}$:
\beq
Q_3, Q_5 =
\begin{cases}
    {\bf 3^*}^a_{I = \frac{1}{2}} + \sqrt{2}\cdot {\bf 3^*}^s_{I = \frac{1}{2}} ~, & q = d ~,\\
    {\bf 3^*}^a_{I = 0} + \sqrt{2}\cdot {\bf 3^*}^s_{I = 0} ~, & q = s~,
\end{cases}
~,~~
Q_4, Q_6 =
\begin{cases}
    -{\bf 3^*}^a_{I = \frac{1}{2}} +\sqrt{2} \cdot {\bf 3^*}^s_{I = \frac{1}{2}} ~, & q=d ~,\\
    -{\bf 3^*}^a_{I = 0} + \sqrt{2} \cdot {\bf3^*}^s_{I = 0} ~, & q=s ~.
\end{cases}
\label{eq:penguin_ops}
\eeq
Finally, the electroweak penguin operators $Q_9$ and $Q_{10}$ [Eq.~(\ref{eq:EWP_oper})] transform as
\beq
Q_9 =
\begin{cases}
\frac32 \left[ - \frac{1}{2} \cdot {\bf 6}_{I = \frac{1}{2}} - \frac{1}{\sqrt{3}}\cdot{\bf 15^*}_{I = \frac{3}{2}} - \frac{1}{2\sqrt{6}}\cdot{\bf 15^*}_{I = \frac{1}{2}} \right] + \frac14 \left( {\bf 3^*}^a_{I = \frac{1}{2}} - \frac{1}{\sqrt{2}} \cdot {\bf 3^*}^s_{I = \frac{1}{2}} \right) ~~, & q=d ~, \\
\frac32 \left[ \frac{1}{2}\cdot {\bf 6}_{I = 1} - \frac{1}{2} \cdot {\bf 15^*}_{I = 1} - \frac{1}{2\sqrt{2}}\cdot{\bf 15^*}_{I = 0} \right] + \frac14 \left({\bf 3^*}^a_{I = 0}- \frac{1}{\sqrt{2}}\cdot{\bf 3^*}^s_{I = 0} \right) ~~, & q=s ~,
\end{cases} \label{EWPopsSU3F}
\eeq
\beq
Q_{10} =
\begin{cases}
\frac32 \left[ \frac{1}{2} \cdot {\bf 6}_{I = \frac{1}{2}} - \frac{1}{\sqrt{3}}\cdot{\bf 15^*}_{I = \frac{3}{2}} - \frac{1}{2\sqrt{6}}\cdot{\bf 15^*}_{I = \frac{1}{2}} \right] + \frac14 \left( -{\bf 3^*}^a_{I = \frac{1}{2}} - \frac{1}{\sqrt{2}} \cdot {\bf 3^*}^s_{I = \frac{1}{2}} \right) ~~, & q=d ~, \\
\frac32 \left[ -\frac{1}{2} \cdot {\bf 6}_{I = 1} - \frac{1}{2} \cdot {\bf 15^*}_{I = 1} - \frac{1}{2\sqrt{2}}\cdot{\bf 15^*}_{I = 0} \right]  + \frac14 \left( -{\bf 3^*}^a_{I = 0} - \frac{1}{\sqrt{2}} \cdot {\bf 3^*}^s_{I = 0} \right) ~~, & q=s ~.
\end{cases} \nn
\eeq

What do we learn from the above expressions? First, compare $Q_1$ and $Q_2$. For both $q=d$ and $q=s$, we see that the coefficients of the ${\bf 15^*}$ and ${\bf 3^*}^s$ terms are the same for $Q_1$ and $Q_2$, while for ${\bf 6}$ and ${\bf 3^*}^a$ they are of opposite sign. This means that, when we consider the full weak Hamiltonian $H_W$, which contains $c_1 Q_1 + c_2 Q_2$, the coefficients of the ${\bf 15^*}$ and ${\bf 3^*}^s$ terms will include $c_1 + c_2$, while those of ${\bf 6}$ and ${\bf 3^*}^a$ will have $c_1 - c_2$. It is similar for $Q_9$ and $Q_{10}$: when one uses $H_W$, the coefficients of the ${\bf 15^*}$ and ${\bf 3^*}^s$ terms will include $c_9 + c_{10}$, while those of the ${\bf 6}$ and ${\bf 3^*}^a$ will have $c_9 - c_{10}$.

Second, according to Eq.~(\ref{EWPtreeops}), $Q_{9} = \frac32 Q_2 - \frac12 Q_3$ and $Q_{10} = \frac32 Q_1 - \frac12 Q_4$. It is straightforward to verify that Eqs.(\ref{TopsSU3F})-(\ref{EWPopsSU3F}) satisfy these relations. Comparing $Q_1$ and $Q_{10}$ in Eqs.~(\ref{TopsSU3F}) and (\ref{EWPopsSU3F}), we see that, apart from the multiplicative factor $\frac32$, $Q_1$ and $Q_{10}$ contain the same ${\bf 6}$ and ${\bf 15^*}$ terms; they differ only in the ${\bf 3^*}$ terms (this is due to the contribution of $Q_4$ to $Q_{10}$). Similarly, the comparison of $Q_2$ and $Q_9$ shows that they contain the same ${\bf 6}$ and ${\bf 15^*}$ terms, but the ${\bf 3^*}$ terms are different.

Because gluonic penguin operators do not contribute to $H_W({\bf 6})$ or $H_W({\bf 15^*)}$, we deduce that the EWP-tree relations arise only from the RMEs involving these operators. There are three such RMEs [see Eq.~(\ref{5RMEs})]. Referring to Eqs.~(\ref{eq:HW})-(\ref{eq:EWP_oper}), we write $H_W = \lambda_u^{(q)} Q_T - \lambda_t^{(q)} (Q_P + Q_{EW})$, where $Q_T = c_1 Q_1 + c_2 Q_2$, $Q_P = \sum_{i=3}^6 c_i Q_i$ and $Q_{EW} = c_9 Q_9 + c_{10} Q_{10}$ (recall that we have neglected the operators $Q_7$ and $Q_8$). The three RMEs of interest can all be written as the sum of two terms, one involving $Q_T$, the other $Q_{EW}$:
\bea
& \langle {\bf 8} || H_W({\bf 6}) || {\bf 3} \rangle = \lambda_u^{(q)}R_8^u - \lambda_t^{(q)}R_8^t ~, & \nn\\
& \langle {\bf 8} || H_W({\bf 15^*}) || {\bf 3} \rangle = \lambda_u^{(q)}P_8^u - \lambda_t^{(q)}P_8^t ~~,~~~
\langle {\bf 27} || H_W({\bf 15^*}) || {\bf 3} \rangle = \lambda_u^{(q)}P_{27}^u - \lambda_t^{(q)}P_{27}^t ~, &
\eea
where
\bea
& R_8^u = \langle {\bf 8} || Q_T({\bf 6}) || {\bf 3} \rangle ~~,~~~~
R_8^t = \langle {\bf 8} || Q_{EW}({\bf 6}) || {\bf 3} \rangle ~, & \nn\\
& P_8^u = \langle {\bf 8} || Q_T({\bf 15^*}) || {\bf 3} \rangle ~~,~~~~
P_8^t = \langle {\bf 8} || Q_{EW}({\bf 15^*}) || {\bf 3} \rangle ~, & \\
& P_{27}^u = \langle {\bf 27} || Q_T({\bf 15^*}) || {\bf 3} \rangle ~~,~~~~
P_{27}^t = \langle {\bf 27} || Q_{EW}({\bf 15^*}) || {\bf 3} \rangle ~. & \nn
\eea

Consider now $R_8^u$ and $R_8^t$. From Eqs.~(\ref{TopsSU3F}) and (\ref{EWPopsSU3F}), for $q=d(s)$, we have
\beq
R_8^u = (c_1 - c_2) \langle {\bf 8} || \frac{1}{2} \cdot {\bf 6}_{I = \frac{1}{2}(1)} || {\bf 3} \rangle
~~,~~~~
R_8^t = \frac32 (c_9 - c_{10}) \langle {\bf 8} || \frac{1}{2} \cdot {\bf 6}_{I = \frac{1}{2}(1)} || {\bf 3} \rangle ~.
\label{eq:R8}
\eeq
This leads to
\beq
R_8^t = \frac32 \, \frac{(c_9 - c_{10})}{(c_1 - c_2)} \, R_8^u ~.
\label{R8ETR}
\eeq
Similar exercises can be carried out for $P_8^{u,t}$ and $P_{27}^{u,t}$, leading to
\bea
P_8^u &=&(c_1 + c_2) \langle {\bf 8} || -\frac{\sqrt{3}}{2\sqrt{2}} \left( \frac{2\sqrt{2}}{3}{\bf 15^*}_{I=\frac{3}{2}(1)} + \frac{1}{3} \cdot {\bf 15^*}_{I=\frac{1}{2}(0)} \right) || {\bf 3} \rangle ~~, \nn \\
P_8^t  &=& \frac{3}{2} (c_9 + c_{10}) \langle {\bf 8} || -\frac{\sqrt{3}}{2\sqrt{2}} \left( \frac{2\sqrt{2}}{3}{\bf 15^*}_{I=\frac{3}{2}(1)} + \frac{1}{3} \cdot {\bf 15^*}_{I=\frac{1}{2}(0)} \right) || {\bf 3} \rangle ~~,
\label{eq:P8}
\eea
\bea
P_{27}^u &=&(c_1 + c_2) \langle {\bf 27} || -\frac{\sqrt{3}}{2\sqrt{2}} \left( \frac{2\sqrt{2}}{3}{\bf 15^*}_{I=\frac{3}{2}(1)} + \frac{1}{3} \cdot {\bf 15^*}_{I=\frac{1}{2}(0)} \right) || {\bf 3} \rangle ~~, \nn \\
P_{27}^t  &=& \frac{3}{2} (c_9 + c_{10}) \langle {\bf 27} || -\frac{\sqrt{3}}{2\sqrt{2}} \left( \frac{2\sqrt{2}}{3}{\bf 15^*}_{I=\frac{3}{2}(1)} + \frac{1}{3} \cdot {\bf 15^*}_{I=\frac{1}{2}(0)} \right) || {\bf 3} \rangle ~~.
\label{eq:P27}
\eea
Examining Eqs.~(\ref{eq:P8}) and (\ref{eq:P27}), we find the following relations:
\beq
P_8^t = \frac32 \, \frac{(c_9 + c_{10})}{(c_1 + c_2)} \, P_8^u ~~,~~~~
P_{27}^t = \frac32 \, \frac{(c_9 + c_{10})}{(c_1 + c_2)} \, P_{27}^u ~.
\label{P8P27ETRs}
\eeq
Eqs.~(\ref{R8ETR}) and (\ref{P8P27ETRs}) are the EWP-tree relations at the level of RMEs.

These can be converted into diagrammatic EWP-tree relations by writing the RMEs as functions of diagrams. In the $\lambda_{u}^{(q)}$ sector, there are five RMEs (two RMEs involving $H_W({\bf 3^*})$ and $R_8^u$, $P_8^u$, $P_{27}^u$) and six diagrams ($T$, $C$, $E$, $A$, $P_{uc}$,  $PA_{uc}$). Only five linear combinations of diagrams -- those equivalent to the RMEs -- appear in the $B \to PP$ amplitudes. For the three RMEs of interest, these combinations of diagrams are
\beq
\begin{split}
R_8^u = -\frac{\sqrt{5}}{4}\left(T -C -A +E\right) &~~,~~~~
P_8^u = \frac{1}{2\sqrt{6}}\left(T + C + 5A + 5E\right) ~, \\
P_{27}^u &= -\sqrt{\frac{2}{3}}\left(T + C\right) ~.
\end{split}
\label{RMEs_u-Diagrams}
\eeq

In the $\lambda_{t}^{(q)}$ sector, there are again five RMEs (two RMEs involving $H_W({\bf 3^*})$ and $R_8^t$, $P_8^t$, $P_{27}^t$). In Ref.~\cite{Gronau:1998fn}, it was shown that there is an EWP diagram associated with each of the six $\lambda_{u}^{(q)}$ diagrams. This means that there are now eight $\lambda_{t}^{(q)}$ diagrams: $P_{tc}$,  $PA_{tc}$ and the six EWP diagrams. Once again, only five linear combinations of the eight diagrams appear in the $B\to PP$ amplitudes. $R_8^t$, $P_8^t$ and $P_{27}^t$ can be expressed as functions of these diagrams:
\bea
R_8^t &=& -\frac{\sqrt5}{4}\left(P_{EW} - P_{EW}^C - P_{EW}^A + P_{EW}^E \right) ~, \nn\\
P_8^t &=& \frac{1}{2\sqrt{6}}\left(P_{EW} + P_{EW}^C + 5 P_{EW}^A + 5 P_{EW}^E \right) ~, \nn\\
P_{27}^t &=& -\sqrt{\frac{2}{3}}\left(P_{EW} + P_{EW}^C\right) ~.
\label{RMEs_t-Diagrams}
\eea

When one inserts Eqs.~(\ref{RMEs_u-Diagrams}) and (\ref{RMEs_t-Diagrams}) into Eqs.~(\ref{R8ETR}) and (\ref{P8P27ETRs}) and solves for the EWP diagrams, the following three relations are found:
\begin{equation}
\begin{split}
    P_{EW} + P_{EW}^E &= -\frac{3}{4}\left[\frac{(c_9 + c_{10})}{(c_1 + c_2)}\left(T+C+A+E\right)+\frac{(c_9 - c_{10})}{(c_1 - c_2)} \left(T-C-A+E\right)\right]\\
    P_{EW}^C - P_{EW}^E &= -\frac{3}{4}\left[\frac{(c_9 + c_{10})}{(c_1 + c_2)}\left(T+C-A-E\right)-\frac{(c_9 - c_{10})}{(c_1 - c_2)} \left(T-C-A+E\right)\right]\\
    P_{EW}^A + P_{EW}^E  &= -\frac{3}{2}\frac{(c_9 + c_{10})}{(c_1 + c_2)} (A+E)
\end{split}
\label{eq:su3_ETR_relations}
\end{equation}
These are the diagrammatic EWP-tree relations. The overall minus sign on the right-hand side of these relations arises from the minus sign associated with the $\lambda_t$ pieces of the Hamiltonian.

We note in passing that the first two EWP-tree relations above can be added together to produce the following EWP-tree relation:
\beq
P_{EW} + P_{EW}^C = -\frac32 \frac{(c_9 + c_{10})}{(c_1 + c_2)} (T + C) ~.
\label{ETR_orig}
\eeq
This ``combined'' EWP-tree relation was first derived in Refs.~\cite{Neubert:1998pt, Neubert:1998jq}, and some analyses \cite{Fleischer:2008wb, Fleischer:2018bld} use only it. As we will see, this relation will reappear below in the sections on SU(2)$_I$ EWP-tree relations.

With these EWP-tree relations, we can now quantify the effect of neglecting the $Q_{7,8}$ operators. These operators contribute to the same RMEs as $Q_{9,10}$. However, since $Q_{7,8}$ are $(V-A) \times (V+A)$ operators, while $Q_{9,10}$ are $(V-A) \times (V-A)$ [see Eq.~(\ref{eq:EWP_oper})], their matrix elements are different. Including the contributions of $Q_{7,8}$, the above combined EWP-tree relation is modified to
\bea \label{eq:ETRmodified}
P_{EW} + P_{EW}^C &=& -\frac32 \frac{((c_9 + c_{10}) + (c_7 + c_8) X)}{(c_1 + c_2)} (T + C) \nn\\
&=& -\frac32 \frac{(c_9 + c_{10})}{(c_1 + c_2)} \left[ 1 + \frac{(c_7 + c_8)}{(c_9 + c_{10})} X \right] (T + C) ~,
\eea
where $X$ is the ratio of $(V-A) \times (V+A)$ and $(V-A) \times (V-A)$ matrix elements. Now, $(c_7 + c_8)/(c_9 + c_{10}) \simeq 10\%$ \cite{Buchalla:1995vs}. Both matrix elements have been calculated for $\btopik$ decays within QCD factorization and $X = O(1)$ is found \cite{Beneke:2001ev}. Assuming this holds for all $B\rightarrow PP$ decays, the above correction is therefore $\simeq 10\%$. (For the terms in the EWP-tree relations of Eq.~(\ref{eq:su3_ETR_relations}) involving $(c_9 - c_{10})/(c_1 - c_2)$, the correction is proportional to $(c_7 - c_8)/(c_9 - c_{10})$, but this is tiny, $\simeq 0.2\%$.) The point is: had $Q_{7,8}$ not been neglected, the EWP-tree relations would change by at most $\simeq 10\%$.

Recently, within the context of SU(3)$_F$, two analyses of $B \to PP$ decays arrived at quite different conclusions. First, as noted above, in Refs.~\cite{Berthiaume:2023kmp,Bhattacharya:2025wcq}, it was found that there is a significant ($4.1\sigma$) discrepancy with the SU(3)$_F$ limit of the SM, and 1000\% SU(3)$_F$ breaking is required to explain the data. On the other hand, in Ref.~\cite{BurgosMarcos:2025xja}, a poor fit is found in the SU(3)$_F$ limit (but is still quite a bit better than that found in Refs.~\cite{Berthiaume:2023kmp,Bhattacharya:2025wcq}), and the addition of factorizable SU(3)$_F$-breaking corrections is sufficient to produce a good fit. It is the different treatments of the EWP-tree relations that explain these different results.

In Refs.~\cite{Berthiaume:2023kmp,Bhattacharya:2025wcq}, the SU(3)$_F$ EWP-tree relations of Eq.~(\ref{eq:su3_ETR_relations}) are imposed, reducing the number of independent RMEs from ten to seven. In contrast, the analysis of Ref.~\cite{BurgosMarcos:2025xja} uses the formalism of Ref.~\cite{He:2018php}. (Earlier analyses of $B \to PP$ decays using this formalism (or a similar one) can be found in Refs.~\cite{Hsiao:2015iiu, Huber:2021cgk}.) In this formalism, the $B \to PP$ amplitudes can be described in terms of SU(3)$_F$ irreducible representations or topological diagrams, and the equivalence of the two descriptions is shown in Ref.~\cite{He:2018php}. But the key point is that EWP-tree relations are not imposed, so there are ten unknown theoretical parameters in the fits. Because there is a larger number of free parameters to vary, Ref.~\cite{BurgosMarcos:2025xja} finds a better agreement with the data than do Refs.~\cite{Berthiaume:2023kmp,Bhattacharya:2025wcq}.

This was checked in Ref.~\cite{Bhattacharya:2025wcq}: the fits were redone allowing the EWPs to vary freely, i.e., the EWP-tree relations were not imposed. The fit was found to improve quite a bit, to the level found in Ref.~\cite{BurgosMarcos:2025xja} (since the observables used in the two fits were a bit different, the values of $\chi^{2}_{\rm min}$ did not match exactly). However, the EWP-tree relations were badly broken: the best-fit values of $P_{EW}$ and $P_{EW}^C$ differed by factors of about 50(!) from the values found using Eq.~(\ref{eq:su3_ETR_relations}). For this to happen, the matrix elements associated with $Q_{7,8}$ must be orders of magnitude larger than those associated with $Q_{9,10}$. If the result of Ref.~\cite{BurgosMarcos:2025xja} is to be believed, it must be demonstrated that such a large enhancement of these matrix elements is possible. (We note in passing that the formalism of Ref.~\cite{He:2018php} was recently updated to include the EWP-tree relations \cite{Shi:2025}.)

\section{\boldmath EWP-tree relations in SU(2)$_I$: $\Delta S = 0$ decays}
\label{section:EWP-tree su2 deltaS0}

The method used in the previous section to derive the EWP-tree relations in the context of SU(3)$_F$ can be adapted straightforwardly to the case where the symmetry is isospin. Under isospin, ($u,d$) and ($\bar{d},-\bar{u}$) form doublets, while $s$ is a singlet. Because of this, $\Delta S=0$ and $\Delta S=1$ decays must be treated separately. In this section, we focus on $\Delta S=0$ decays.

The weak Hamiltonian that governs $\Delta S = 0$ decays is the same as that used in $SU(3)$ [Eq. (\ref{eq:HW})], for $q=d$. At the quark level, the decay involves the transitions ${\bar b} \to {\bar d} u {\bar u}$ and ${\bar b} \to ({\bar q} d {\bar d}+{\bar q} s {\bar s})$, so the weak Hamiltonian transforms under SU(2)$_I$ as ${\cal O}_{1/2}^{3/2}$ or ${\cal O}_{1/2}^{1/2}$ (in ${\cal O}^I_{I_3}$, $I$ is the isospin and $I_3$ is the third component of isospin).

The individual operators can be written in terms of these SU(2)$_I$ representations. For the tree operators, we have
\beq
Q_1 =
-\frac{1}{\sqrt3}{\cal O}_{1/2}^{3/2} + \frac{1}{\sqrt6}{\cal {O}}_{1/2}^{1/2} + \frac{1}{\s} {\cal {O}}_{1/2}^{'1/2}~~,~~~~
Q_2 =
-\frac{1}{\sqrt3}{\cal O}_{1/2}^{3/2} + \frac{1}{\sqrt6}{\cal {O}}_{1/2}^{1/2} - \frac{1}{\s} {\cal {O}}_{1/2}^{'1/2} ~.
\label{TopsSU2dS0}
\eeq
The gluonic penguin operators transform as a doublet under $SU(2)_I$. For instance, $Q_3$ and $Q_4$ transform as
\beq
Q_3 = \frac{\sqrt3}{\s}{\cal O}_{1/2}^{1/2}-\frac{1}{\s}{\cal O}_{1/2}^{'1/2}+{\cal O}_{1/2}^{''1/2}~~,~~~~
Q_4 = \frac{\sqrt3}{\s}{\cal O}_{1/2}^{1/2}+\frac{1}{\s}{\cal O}_{1/2}^{'1/2}+{\cal O}_{1/2}^{''1/2} ~.
\label{eq:penguin_ops_su2dS0}
\eeq
Finally, the electroweak penguin operators transform under isospin as
\beq
Q_9  = -\frac{\sqrt3}{2}{\cal O}_{1/2}^{3/2} -\frac{1}{\s}{\cal O}_{1/2}^{'1/2}-\frac{1}{2}{\cal O}_{1/2}^{''1/2} ~~,~~~~
Q_{10} = -\frac{\sqrt3}{2}{\cal O}_{1/2}^{3/2} +\frac{1}{\s}{\cal O}_{1/2}^{'1/2}-\frac{1}{2}{\cal O}_{1/2}^{''1/2} ~.
\label{eq:EWP_ops_su2dS0}
\eeq
It is straightforward to verify that Eqs.~(\ref{TopsSU2dS0}), (\ref{eq:penguin_ops_su2dS0}) and (\ref{eq:EWP_ops_su2dS0}) satisfy Eq.~(\ref{EWPtreeops}), the relations expressing the EWP operators in terms of the tree and gluonic penguin operators.

From Eq.~(\ref{TopsSU2dS0}), we see that the coefficients of the ${\cal O}_{1/2}^{3/2}$ and ${\cal {O}}_{1/2}^{1/2}$ terms are the same in both $Q_1$ and $Q_2$, so these operators in the full Hamiltonian will include $(c_1+c_2)$. On the other hand, the factor in front of ${\cal {O}}_{1/2}^{'1/2}$ in $Q_1$ is equal, but of opposite sign, to the one in $Q_2$, so its coefficient in the full Hamiltonian will include $(c_1-c_2)$. Similarly, from Eq.~(\ref{eq:EWP_ops_su2dS0}), which focuses on $Q_9$ and $Q_{10}$, we see that the coefficients of ${\cal O}_{1/2}^{3/2}$ and ${\cal O}_{1/2}^{''1/2}$ in the full Hamiltonian will include $(c_9+c_{10})$, while that of ${\cal O}_{1/2}^{'1/2}$ will include $(c_9-c_{10})$.

For the initial state, $B = (B^+, B^0)$ forms a doublet under isospin, while $B^0_s$ is a singlet. For $\Delta S =0$ decays, we can separate the eight decay modes into three subsets based on the isospin of the initial and final states. These subsets are $B\to\pi\pi$ (three decays),
$\bs\to\pi {\bar K}$ (two decays) and $B\to K {\bar K}$ (three decays). Below, we derive the EWP-tree relations for each of these subsets. As was the case in the SU(3)$_F$ analysis, EWP-tree relations arise from those RMEs to which gluonic penguin operators do not contribute. We therefore focus only on those RMEs that involve the SU(2)$_I$ operator ${\cal O}_{1/2}^{3/2}$.

\subsection{\bf\boldmath $B\to\pi\pi$}
\label{Bpipi}

The three $B \to\pi\pi$ decays are $B^+\to\pi^+ \pi^0$, $B^0\to\pi^+\pi^-$ and $B^0\to\pi^0\pi^0$. The initial state $B = (B^+, B^0)$ has isospin $\frac12$, while the final $\pi\pi$ states can have isospin 0 or 2. There are, therefore, only two RMEs that contribute to these decays. They are $\langle 0 || H_W^{1/2} || \frac12 \rangle$ and $\langle 2 || H_W^{3/2} || \frac12 \rangle$.

For the RME containing $H_W^{3/2}$, we write
\bea
\langle 2 || H_W^{3/2} || \frac12 \rangle = \la_u^{(d)}A^u_{3/2,2} - \la_t^{(d)}A^t_{3/2,2}~,
\label{eq:Hw_BpipiDS0}
\eea
where, from Eqs.~(\ref{TopsSU2dS0}) and (\ref{eq:EWP_ops_su2dS0}),
\bea
A^u_{3/2,2} =\langle 2 || Q_T^{3/2} || \frac12 \rangle = ~~~(c_1+c_2)\langle 2 || -\frac{1}{\sqrt3} {\cal O}_{1/2}^{3/2}|| \frac12 \rangle ~,\nn\\~~~~
A^t_{3/2,2} = \langle 2 || Q_{EW}^{3/2} || \frac12 \rangle = \frac32 (c_9+c_{10})\langle 2 || -\frac{1}{\sqrt3} {\cal O}_{1/2}^{3/2}|| \frac12 \rangle~.
\label{preETR}
\eea
Inserting these into Eq.~(\ref{preETR}), we obtain an EWP-tree relation at the level of RMEs:
\bea
A^t_{3/2,2} = \frac32 \frac{(c_9+c_{10})}{(c_1+c_2)}A^u_{3/2,2} ~.
\label{BpipiETR}
\eea

This can be converted into a diagrammatic EWP-tree relation by writing the RMEs as functions of the diagrams contributing to $B \to \pi\pi$ decays. In the $\la_u^{(d)}$ sector, we have
\beq
A^u_{3/2,2} = -\sqrt{\frac23}(T + C) ~,
\eeq
and in the $\la_t^{(d)}$ sector, we have
\beq
A^t_{3/2,2} = -\sqrt{\frac23} (P_{EW} + P_{EW}^C) ~.
\eeq
These lead to the following EWP-tree relation at the level of diagrams:
\beq
P_{EW} + P_{EW}^C = -\frac32 \frac{(c_9 + c_{10})}{(c_1 + c_2)} (T + C) ~.
\label{ETR_isospin}
\eeq

At first sight, this is identical to the combined SU(3)$_F$ EWP-tree relation of Eq.~(\ref{ETR_orig}). However, there are nuances. First, although the SU(3)$_F$ and SU(2)$_I$ diagrams have the same symbols, they are not really identical. In SU(3)$_F$ the $T$ diagram (for example) is the same for  many different decays, whereas in SU(2)$_I$ it applies only to $B \to\pi\pi$ decays -- the $T$ diagrams in other decays are different. Second, SU(3)$_F$ is broken in the SM at the level of $\sim 25\%$, whereas isospin is an almost exact symmetry. Therefore, even though they look the same, we expect the SU(2)$_I$ EWP-tree relation of Eq.~(\ref{ETR_isospin}) to hold almost exactly, while the SU(3)$_F$ EWP-tree relation of Eq.~(\ref{ETR_orig}) has a $\sim 25\%$ uncertainty.

\subsection{\bf\boldmath $\bs\to\pi {\bar K}$}

The two decays in this subset are $\bs\to\pi^+ K^-$ and $\bs\to\pi^0 \bar K^0$.
The final states have isospin $\frac12$ or $\frac32$, while the initial state $\bs$ is a singlet of SU(2)$_I$ and has isospin 0. The two RMEs that contribute to these decays are $\langle \frac12 || H_W^{1/2} || 0 \rangle$ and $\langle \frac32 || H_W^{3/2} || 0 \rangle$. Since EWP-tree relations arise only from RMEs involving $H_W^{3/2}$, we focus only on the second RME, writing
\beq
\langle \frac32 || H_W^{3/2} || 0 \rangle = \la_u^{(d)}A^u_{3/2,3/2} - \la_t^{(d)}A^t_{3/2,3/2} ~,
\label{eq:Hw_BpiKDS0}\eeq
where [similar to Eq.~(\ref{preETR})]
\bea
A^u_{3/2,3/2} = \langle \frac32 || Q_T^{3/2} || 0 \rangle =~~~(c_1+c_2)\langle \frac32 || -\frac{1}{\sqrt3} {\cal O}_{1/2}^{3/2}|| 0 \rangle~,\nonumber\\
A^t_{3/2,3/2} = \langle \frac32 || Q_{EW}^{3/2} || 0 \rangle = \frac32(c_9+c_{10})\langle \frac32 || -\frac{1}{\sqrt3} {\cal O}_{1/2}^{3/2}|| 0 \rangle~.
\eea
This leads to the following EWP-tree relation at the RME level:
\beq
A^t_{3/2,3/2} = \frac32 \frac{(c_9+c_{10})}{(c_1+c_2)}A^u_{3/2,3/2}~.
\eeq

For $\bs\to\pi {\bar K}$ decays, the relations between RMEs and diagrams in the $\la_u^{(d)}$ and $\la_t^{(d)}$ sectors are
\beq
\la_u^{(d)} ~:~~
A^u_{3/2,3/2} ~=~ -\frac{1}{\sqrt3}(T +C) ~~,~~~~
\la_t^{(d)} ~:~~ A^t_{3/2,3/2} ~=~ -\frac{1}{\sqrt3}(P_{EW} + P_{EW}^C) ~.
\eeq
These produce the following diagrammatic EWP-tree relation:
\beq
P_{EW} + P_{EW}^C =-\frac32\frac{c_9+c_{10}}{c_1+c_2}(T +C) ~.
\label{ETR_isospin2}
\eeq
As was the case for $B \to\pi\pi$ decays,
this is just Eq.~(\ref{ETR_orig}), one of the SU(3)$_F$ EWP-tree relations.

\subsection{\bf\boldmath $B\to K{\bar K}$}

There are three decays in this subset: $B^+\to K^+ \bar K^0$, $B^0\to K^0\bar K^0$ and $B^0\to K^+ K^-$. The final states have isospin 0 or 1, while the initial state $B = (B^+, B^0)$ has isospin $\frac12$. There are three RMEs that contribute to these decays: $\langle 0 || H_W^{1/2} || \frac12 \rangle$, $\langle 1 || H_W^{1/2} || \frac12 \rangle$ and $\langle 1 || H_W^{3/2} || \frac12 \rangle$. Focusing only on the RME that contains $H_W^{3/2}$, we write
\beq
\langle 1 || H_W^{3/2} || \frac12 \rangle = \la_u^{(d)}A^u_{3/2,1} - \la_t^{(d)}A^t_{3/2,1} ~,
\label{eq:Hw_BKKDS0}\eeq
where [similar to Eq.~(\ref{preETR})]
\bea
A^u_{3/2,1} = \langle 1 || Q_T^{3/2} || \frac12 \rangle = ~~ ~(c_1+c_2)\langle 1 || -\frac{1}{\sqrt3} {\cal O}_{1/2}^{3/2}|| \frac12 \rangle~,\nonumber\\
A^t_{3/2,1} = \langle 1 || Q_{EW}^{3/2} || \frac12 \rangle = \frac32(c_9+c_{10})\langle 1 || -\frac{1}{\sqrt3} {\cal O}_{1/2}^{3/2}|| \frac12 \rangle\textbf{} ~.
\eea
This leads to the following EWP-tree relation at the level of RMEs:
\beq
A^t_{3/2,1} = \frac32 \frac{(c_9+c_{10})}{(c_1+c_2)}A^u_{3/2,1} ~.
\eeq

For $B\to K{\bar K}$ decays, the relations between RMEs and diagrams are
\beq
\la_u^{(d)} ~:~~ A^u_{3/2,1} = -\frac23 (A + E) ~~,~~~~
\la_t^{(d)} ~:~~
A^t_{3/2,1} = -\frac23 (P_{EW}^A + P_{EW}^E) ~.
\eeq
This leads to an EWP-tree relation at the level of diagrams:
\beq
P_{EW}^A + P_{EW}^E =-\frac32\frac{c_9+c_{10}}{c_1+c_2} (A + E) ~.
\eeq
This has the same form as the third SU(3)$_F$ EWP-tree relation of Eq.~(\ref{eq:su3_ETR_relations}).

\section{\boldmath EWP-tree relations in SU(2)$_I$: $\Delta S =1$ decays}
\label{section:EWP-tree su2 deltaS1}

At the quark level, $\Delta S=1$ decays involve the transitions ${\bar b} \to {\bar s} u {\bar u}$ or ${\bar b} \to {\bar s}$. The weak Hamiltonian therefore transforms under SU(2)$_I$ as an ${\cal O}^1_0$ or an ${\cal O}^0_0$.

We write the operators of the weak Hamiltonian in terms of these SU(2)$_I$ representations. For the tree operators $Q_1$ and $Q_2$, we have
\begin{equation}
Q_1 = Q_2 =\frac{1}{\sqrt{2}} (\mathcal{O}_0^0- \mathcal{O}_0^1) ~.
\label{eq:tree_ops_su2dS1}
\end{equation}
The gluonic penguin operators are purely ${\bar b} \to {\bar s}$ transitions, i.e., they are singlets under SU(2)$_I$. In particular, $Q_3$ and $Q_4$ transform as
\beq
Q_3 = Q_4 = \sqrt{2} {\cal O}_{0}^{0}+{\cal O}_{0}^{'0} ~.
\label{eq:penguin_ops_su2dS1}
\eeq
Finally, the EWP operators $Q_9$ and $Q_{10}$ transform as
\begin{equation}
 Q_9 = Q_{10} =     \frac{1}{2\sqrt{2}} (\mathcal{O}_0^0- 3\mathcal{O}_0^1) -\frac{1}{2}{\cal O}_{0}^{'0} ~.
 \label{eq:EWP_ops_su2dS1}
\end{equation}
It is straightforward to verify that Eqs.~(\ref{eq:tree_ops_su2dS1}), (\ref{eq:penguin_ops_su2dS1}) and (\ref{eq:EWP_ops_su2dS1}) satisfy Eq.~(\ref{EWPtreeops}).

From Eq.~(\ref{eq:tree_ops_su2dS1}) (Eq.~(\ref{eq:EWP_ops_su2dS1})), we see that $Q_1$ and $Q_2$ ($Q_9$ and $Q_{10}$) transform identically under SU(2)$_I$. As a result, when the full Hamiltonian is used, it will only involve the combination $c_1+c_2$ ($c_9+c_{10}$). This is in contrast to the calculation with SU(3)$_F$ or with SU(2)$_I$ ($\Delta S=0$), where both combinations $c_1 \pm c_2$ ($c_9 \pm c_{10}$) appeared.

As was the case with $\Delta S=0$ decays, the eight $\Delta S=1$ decay modes can be separated into three subsets. They are $B \to \pi K$ (four decays), $\bs \to K{\bar K}$ (two decays) and $\bs \to \pi \pi$ (two decays). In what follows, we derive the EWP-tree relations for each of these subsets. Once again, we focus only on those RMEs that receive no contributions from the gluonic penguin operators. Since the gluonic penguin operators do not contribute to ${\cal O}_{0}^{1}$, in this case, only the RMEs that involve the operator $\mathcal{O}_0^1$ are important.

\subsection{\bf\boldmath $\btopik$}
\label{BpiKsection}

\subsubsection{\bf EWP-tree relations}

There are four $\btopik$ decay modes: $B^+ \to \pi^+ K^0$, $B^+ \to \pi^0 K^+$, $B^0 \to \pi^- K^+$ and $B^0 \to \pi^0 K^0$. The final $\pi K$ states have isospin $\frac12$ or $\frac32$, while the initial state $B = (B^+, B^0)$ has isospin $\frac12$. There are three RMEs that describe these decays: $\langle \frac12 || H_W^0 || \frac12 \rangle$, $\langle \frac12 || H_W^1 || \frac12 \rangle$ and $\langle \frac32 || H_W^1 || \frac12 \rangle$. These RMEs can be split into pieces proportional to $\la_u^{(s)}$ and $\la_t^{(s)}$:
\bea
\langle \frac12 || H_W^0 || \frac12 \rangle = \la_u^{(s)}A_{0,1/2}^u -\la_t^{(s)}A_{0,1/2}^t~, \nn\\
\langle \frac12 || H_W^1 || \frac12 \rangle = \la_u^{(s)}A_{1,1/2}^u -\la_t^{(s)}A_{1,1/2}^t~, \\
\langle \frac32 || H_W^1 || \frac12 \rangle = \la_u^{(s)}A_{1,3/2}^u -\la_t^{(s)}A_{1,3/2}^t~. \nn
\eea

Since EWP-tree relations arise only from RMEs involving $\mathcal{O}_0^1$, we focus only on the final two RMEs. From Eqs.~(\ref{eq:tree_ops_su2dS1}) and (\ref{eq:EWP_ops_su2dS1}), we have\footnote{$A^{u(t)}_{0,1/2}$ can be similarly expressed in terms of Wilson coefficients and matrix elements of the ${\cal O}^0_0$ operator. However, such expressions are not relevant for the current discussion.}
\bea
& A_{1,1/2}^u = \langle \frac{1}{2}  || Q_{T}^{1} || \frac{1}{2}  \rangle ~~= ~~(c_1+c_2)\langle \frac{1}{2}  || -\frac{1}{\sqrt{2}}.\mathcal{O}^1_0 || \frac{1}{2}  \rangle ~,\nonumber \\
& A_{1,1/2}^t = \langle \frac{1}{2}  || Q_{EW}^{1} || \frac{1}{2}  \rangle = \frac32(c_9+c_{10})\langle \frac{1}{2}  || -\frac{1}{\sqrt{2}}.\mathcal{O}^1_0 || \frac{1}{2}  \rangle ~, & \nonumber \\
& A_{1,3/2}^u = \langle \frac{3}{2}  || Q_{T}^{1} || \frac{1}{2}  \rangle~~ =~~(c_1+c_2)\langle \frac{3}{2}  || -\frac{1}{\sqrt{2}}.\mathcal{O}^1_0 || \frac{1}{2}  \rangle ~, \nonumber \\
&A_{1,3/2}^t = \langle \frac{3}{2}  || Q_{EW}^{1} || \frac{1}{2}  \rangle = \frac32(c_9+c_{10})\langle \frac{3}{2}  || -\frac{1}{\sqrt{2}}.\mathcal{O}^1_0 || \frac{1}{2}  \rangle ~.
\label{eq:BpiKRME}
\eea
These lead to the following two EWP-tree relations at the level of RMEs:
\bea
A_{1,1/2}^t  = \frac{3}{2} \, \frac{c_9+c_{10}}{c_1+c_2} \, A_{1,1/2}^u, ~~~~~A_{1,3/2}^t  = \frac{3}{2} \, \frac{c_9+c_{10}}{c_1+c_2} \, A_{1,3/2}^u
\label{eq:RMEEWPBpiK}
\eea

For $\btopik$ decays, the relations between the RMEs of interest and diagrams in the $\la_u^{(s)}$ sector are as follows:
\beq
A_{1,1/2}^u = \frac{1}{2\sqrt{2}} ( T - 2 C - 3 A ) ~~,~~~~
A_{1,3/2}^u =-\frac{1}{\sqrt{2}}( T + C ) ~.
 \label{eq:isolambdau}
 \eeq
In the $\la_t^{(s)}$ sector, we have
\beq
A_{1,1/2}^t = -\frac{1}{2\sqrt{2}}( 2 P_{EW}- P_{EW}^C + 3 P_{EW}^E ) ~~,~~~~
A_{1,3/2}^t = -\frac{1}{\sqrt{2}}( P_{EW} + P_{EW}^{C}) ~.
\label{eq:isolambdat}
\eeq
From Eqs.~(\ref{eq:RMEEWPBpiK}), (\ref{eq:isolambdau}) and (\ref{eq:isolambdat}), we find the two diagrammatic EWP-tree relations to be
\beq
P_{EW} + P_{EW}^E = -\frac{3}{2}\frac{c_9+c_{10}}{c_1+c_2} \, (C+A) ~~,~~~~
P_{EW}^C - P_{EW}^E = -\frac{3}{2}\frac{c_9+c_{10}}{c_1+c_2} \, (T-A) ~.
\label{eq:ETRBPiK}
\eeq
We note that Ref.~\cite{Buras:1998rb} contains a relation between $P_{EW}^{C}$ and $T$ similar to the second relation above.

Interestingly, these $\btopik$ SU(2)$_I$ EWP-tree relations are quite different from those found with SU(3)$_F$ [Eq.~(\ref{eq:su3_ETR_relations})]. This is not entirely surprising. Consider the quark-level transitions ${\bar b} \to {\bar d} u {\bar u}$ ($\Delta S=0$) and ${\bar b} \to {\bar s} u {\bar u}$ ($\Delta S=1$). In the context of SU(3)$_F$, the weak Hamiltonian for both decays transforms as the product of three fundamental representations (${\bf 3}$ or ${\bf 3^*}$). Under SU(2)$_I$, the weak Hamiltonian for $\Delta S=0$ decays also transforms as the product of three fundamental representations (${\bf 2}$). However, under SU(2)$_I$, the weak Hamiltonian for $\Delta S=1$ decays transforms as the product of two fundamental representations and a singlet. This suggests (but does not prove) that the SU(2)$_I$ EWP-tree relations for $\Delta S=0$ decays could be similar to those for SU(3)$_F$, while those for $\Delta S=1$ decays could be different.

This said, if we add the above two EWP-tree relations, we recover the previous results of Eq.~(\ref{ETR_orig}) [SU(3)$_F$] and Eqs.~(\ref{ETR_isospin}) and (\ref{ETR_isospin2}) [SU(2)$_I$, $\Delta S=0$].

\subsubsection{\bf\boldmath A relation among $\btopik$ observables (RMEs)}
\label{Rel_observables_RMEs}

The amplitudes of the four $\btopik$ decays, $B^+ \to \pi^+ K^0$, $B^+ \to \pi^0 K^+$, $B^0 \to \pi^- K^+$ and $B^0 \to \pi^0 K^0$, obey an isospin quadrilateral relation \cite{Nir:1991}:
\beq
A^{+0} + \s A^{0+} = A^{-+} + \s A^{00} ~,
\label{eq:isoquad}
\eeq
where the indices $i$ and $j$ in $A^{ij}$ indicate the charges of the final-state $\pi$ and $K$, respectively. These amplitudes can be expressed in terms of (all) the SU(2)$_I$ RMEs as
\bea
A^{-+} &=& \lambda_{u}^{(s)} \[ -\sqrt{\frac{2}{3}} \(A_{0,1/2}^u+\frac{1}{\sqrt{3}} A_{1,1/2}^u \) +\frac{\sqrt{2}}{3} A_{1,3/2}^u\] \nn \\
&&\hspace{1truecm} -~\lambda_{t}^{(s)} \[ -\sqrt{\frac{2}{3}} \(A_{0,1/2}^t+\frac{1}{\sqrt{3}} A_{1,1/2}^t \) +\frac{\sqrt{2}}{3} A_{1,3/2}^t\]  ~, \nn \\
A^{+0} &=& \lambda_{u}^{(s)} \[\sqrt{\frac{2}{3}} \(A_{0,1/2}^u-\frac{1}{\sqrt{3}} A_{1,1/2}^u \) +\frac{\sqrt{2}}{3} A_{1,3/2}^u\] \nn \\
&&\hspace{1truecm} -~\lambda_{t}^{(s)}\[\sqrt{\frac{2}{3}} \(A_{0,1/2}^t-\frac{1}{\sqrt{3}} A_{1,1/2}^t \) +\frac{\sqrt{2}}{3} A_{1,3/2}^t\] ~, \nn \\
\s\,A^{0+} &=& \lambda_{u}^{(s)} \[-\sqrt{\frac{2}{3}} \(A_{0,1/2}^u -\frac{1}{\sqrt{3}} A_{1,1/2}^u\) +\frac{2\sqrt{2}}{3} A_{1,3/2}^u \] \nn \\
&&\hspace{1truecm} -~\lambda_{t}^{(s)}\[-\sqrt{\frac{2}{3}} \(A_{0,1/2}^t -\frac{1}{\sqrt{3}} A_{1,1/2}^t\) +\frac{2\sqrt{2}}{3} A_{1,3/2}^t \] ~,
\label{ampsSU(2)}
\eea
with $A^{00}$ satisfying Eq.~(\ref{eq:isoquad}). The amplitudes for the CP-conjugate processes can be obtained from the above by changing the sign of the weak phases.

Now, for each $\btopik$ mode, one can construct the following observable:
\beq
\Delta^{ij} ~\equiv~ \frac{{\cal B}^{ij}A_{CP}^{ij}}{\Pi^{ij}} ~=~ |{\overline A}^{ij}|^2 - |A^{ij}|^2\,,
\eeq
where ${\cal B}^{ij}, A^{ij}_{CP}$, and $\Pi^{ij}$ are respectively the CP-averaged branching ratio, the direct CP asymmetry, and the two-body phase-space factor for the decay $B \to\pi^iK^j$. One can combine the $\Delta^{ij}$ for the four decays and compute
\beq
\de_{\pi K} ~=~ \Delta^{-+} + \Delta^{+0} - 2\,\Delta^{0+} - 2\,\Delta^{00}\,.
\eeq
In Ref.~\cite{Gronau:2005kz}, it was pointed out that the leading-order terms in $\de_{\pi K}$ cancel within isospin symmetry. It was then argued that, if one assumes SU(3)$_F$ symmetry (i.e., one uses the SU(3)$_F$ EWP-tree relations) and one adds some theory input (assumptions are made about the magnitudes and strong phases of certain diagrams), the subleading terms also cancel, leading to $\de_{\pi K} \approx 0$. This was touted as a test of the SM.
The problem is that theoretical input has been used, and it is difficult to quantify its uncertainty. Thus, although this is categorized as a null test of the SM, it is nevertheless true that  $\de_{\pi K} > 0$ is possible, and it is not clear just how large $\de_{\pi K}$ is allowed to be.

This ambiguity arises because the SU(3)$_F$ EWP-tree relations were used. However, if one instead uses
the $\btopik$ SU(2)$_I$ EWP-tree relations, one can show that $\de_{\pi K} = 0$ is {\it exact}. That is, additional theory input is unnecessary. We can see this by using Eqs.~(\ref{eq:isoquad}) and (\ref{ampsSU(2)}), along with the definitions of $\Delta^{ij}$ and $\de_{\pi K}$ given above. In terms of the SU(2)$_I$ RMEs, we find
\bea
\de_{\pi K} &=& \frac{16}{3}\,{\rm Im}\[\lambda^{(s)}_{u}\lambda^{(s)*}_{t}\]{\rm Im}\[A^{t*}_{1,3/2}A^{u}_{1,1/2} + A^{t*}_{1,1/2}A^{u}_{1,3/2} + A^{t*}_{1,3/2}A^{u}_{1,3/2}\] \nn \\
&=& 8\,\frac{c_9 + c_{10}}{c_1 + c_2}\,{\rm Im}\[\lambda^{(s)}_{u}\lambda^{(s)*}_{t}\]{\rm Im}\[A^{u*}_{1,3/2}A^u_{1,1/2} + A^{u*}_{1,1/2}A^u_{1,3/2} + A^{u*}_{1,3/2}A^u_{1,3/2}\]\,,~~~~~
\eea
where we have applied the SU(2)$_I$ EWP-tree relations of Eq.~(\ref{eq:RMEEWPBpiK}). The expression within the second set of square brackets, which depends on the SU(2)$_I$ RMEs, is clearly purely real. Its imaginary part vanishes, so that $\de_{\pi K} = 0$ is exact under SU(2)$_I$.

This shows that one must be careful in applying the SU(3)$_F$ EWP-tree relations to $\btopik$ decays -- one may obtain misleading results. We will see this again in Sec.~\ref{section:btopik}. More generally, if one is analyzing a set of $B$ decays whose amplitudes are related by isospin, one must use the EWP-tree relations derived using SU(2)$_I$ for that set of decays, rather than SU(3)$_F$.

Suppose now that $\de_{\pi K}$ were measured to be different from zero. What could explain this? A first thought is that, since isospin is close to an exact symmetry of the SM, the measurement of $\de_{\pi K} \ne 0$ would point to isospin-violating NP. However, since such NP is very stringently constrained, it is unlikely to lead to $\de_{\pi K}$ much different from 0. A second possibility is related to the $Q_{7,8}$ operators. If these operators had not been neglected, the EWP-tree relations of Eq.~(\ref{eq:RMEEWPBpiK}) would be modified similarly to Eq.~(\ref{eq:ETRmodified}):
\bea
A_{1,1/2}^t  &=& \frac{3}{2} \, \frac{c_9+c_{10}}{c_1+c_2} \,
\left[ 1 + \frac{(c_7 + c_8)}{(c_9 + c_{10})} X \right] \, A_{1,1/2}^u ~, \nn\\
A_{1,3/2}^t &=& \frac{3}{2} \, \frac{c_9+c_{10}}{c_1+c_2} \,
\left[ 1 + \frac{(c_7 + c_8)}{(c_9 + c_{10})} X \right] \, A_{1,3/2}^u ~.
\eea
In this case, if Im[$X$] were nonzero, it would lead to $\de_{\pi K} \ne 0$. This is potentially a test of our assumption of neglecting the operators $Q_{7,8}$.

In fact, $\de_{\pi K}$ has been measured,  most recently by Belle II \cite{Belle-II:2023ksq}. They find $\delta_{\pi K} = -0.03 \pm 0.13 \pm 0.04$, consistent with zero. This shows that Im[$X$] is close to zero (but says nothing about Re[$X$]).

\subsection{\bf\boldmath $\bs\to K{\bar K}$}

The decay modes considered here are $\bs \to K^{+} K^{-}$ and $\bs \to K^0 \bar{K}^0$. The final states have an isospin of 0 or 1, while the initial state is an SU(2)$_I$ singlet. Thus, there are two RMEs that contribute to these decays, $\langle 0 || H_W^0 || 0 \rangle$ and $\langle 1 || H_W^1 || 0 \rangle$. For the RME containing $H_W^1$, we write
\begin{equation}
\langle 1 || H_W^1 || 0 \rangle = \la_u^{(s)} A_{1,1}^u -\la_t^{(s)}A_{1,1}^t ~,
\end{equation}
where (using Eqs.~(\ref{eq:tree_ops_su2dS1}) and (\ref{eq:EWP_ops_su2dS1}))
\bea
A_{1,1}^u &=& \langle 1  || Q_{T}^{1} || 0  \rangle ~~= ~~~(c_1+c_2)\langle 1  || -\frac{1}{\sqrt{2}}.\mathcal{O}^1_0 || 0  \rangle ~, \nonumber \\
A_{1,1}^t &=& \langle 1  || Q_{EW}^{1} || 0  \rangle = \frac32(c_9+c_{10})\langle 1 || -\frac{1}{\sqrt{2}}.\mathcal{O}^1_0 || 0  \rangle ~.
\eea
This leads to an EWP-tree relation at the level of RMEs:
\beq
A_{1,1}^t  = \frac{3}{2} \frac{c_9+c_{10}}{c_1+c_2} A_{1,1}^u
\label{eq:RMEEWPBsKK}
\eeq

For $\bs\to K{\bar K}$ decays, the relations between RMEs and diagrams are
\beq
\la_u^{(s)} ~:~~ A_{1,1}^u = -\frac{1}{\sqrt{2}} \, (T + E) ~~,~~~~
\la_t^{(s)} ~:~~ A_{1,1}^t = -\frac{1}{\sqrt{2}} \, (P_{EW}^C + P_{EW}^{A}) ~.
\eeq
This leads to the following diagrammatic EWP-tree relation:
\beq
P_{EW}^C + P_{EW}^{A} = - \frac{3}{2} \frac{c_9+c_{10}}{c_1+c_2} \, (T + E ) ~.
\eeq

Similarly to the $\btopik$ subset, this EWP-tree relation is different from those found in SU(3)$_F$. This can be explained by the same arguments given in the previous section.

\subsection{\bf\boldmath $\bs\to\pi\pi$}

There are two decay modes in this subset, $\bs \to \pi^+ \pi^-$ and $\bs \to \pi^0 \pi^0$. The final $\pi\pi$ states can have isospin 0 or 2. Because the initial state is an SU(2)$_I$ singlet, there is only one RME, $\langle 0 || H_W^0 || 0 \rangle$. But we have already found that EWP-tree relations can only arise from RMEs that involve $H_W^1$. As there are no such RMEs, we therefore conclude that, for this set of decays, there is no EWP-tree relation.

\section{\boldmath Extracting $\alpha$ from $B \to\pi\pi$ decays}
\label{Sec:alphaBpipi}

In the previous sections, we showed that, even if only isospin SU(2)$_I$ symmetry is assumed, there are still EWP-tree relations. The question then is: what can they be used for? EWP-tree relations of a certain symmetry are generally used in analyses of decays whose amplitudes are related by this symmetry. For example, as noted in the introduction, the amplitudes for all charmless $B \to PP$ decays are related to one another by flavor SU(3)$_F$ symmetry. Because of this, when global analyses of charmless $B \to PP$ decays were carried out in Refs.~\cite{Berthiaume:2023kmp,Bhattacharya:2025wcq}, the SU(3)$_F$ EWP-tree relations were used.

One set of decays whose amplitudes are related by isospin, and which have been analyzed on their own, is $B \to \pi\pi$. In Ref.~\cite{Gronau:1990ka}, it was shown that the weak, CP-violating phase $\alpha$ can be extracted from the measurements of the $B \to\pi\pi$ observables. At the time, the demonstration was a geometrical construction. However, today we use a more modern technique, namely a fit to all the $B\to\pi\pi$ data. In order to show that the extraction of $\alpha$ is possible, (i) the $B \to\pi\pi$ amplitudes must be expressed in terms of a certain number of unknown theoretical parameters that includes $\alpha$, and (ii) the number of observables must be greater than or equal to the number of unknown parameters. Let us examine these requirements in turn.

The three $B \to\pi\pi$ decays are $B^+\to\pi^+ \pi^0$, $B^0\to\pi^+\pi^-$ and $B^0\to\pi^0\pi^0$. We can count the number of (potentially) measurable observables in these decays. There are three branching ratios and three direct CP asymmetries. One can also measure the indirect (mixing-induced) CP asymmetry in $B^0\to\pi^+\pi^-$, for a total of seven observables\footnote{In principle, $B^0 \to\pi^0\pi^0$ also has an indirect CP asymmetry. However, because each $\pi^0$ decays to two photons, the time dependence of the amplitude cannot be reconstructed accurately enough to measure the indirect CP asymmetry.}. Note that this assumes that the amplitude for $B^+ \to\pi^+\pi^0$ has two contributions with nonzero relative weak and strong phases. If this is not the case, the number of observables is reduced to six.

The amplitudes of the three $B\to\pi\pi$ decays obey an isospin triangle relation:
\beq
\sqrt{2} A^{+0} = A^{+-} + \sqrt{2} A^{00} ~,
\eeq
where the indices $ij$ in $A^{ij}$ indicate the charges of the final-state pions. These amplitudes can be expressed in terms of RMEs or diagrams. Here, we choose diagrams: the decomposition of the $B \to \pi\pi$ decay amplitudes in terms of diagrams is shown in Table \ref{tab:isopsindiagdeltaS0pipi}.

\begin{table}[h]
\centering
\setlength{\tabcolsep}{4pt}
\begin{tabular}{|c|c|c|c|c|c|c|c|c|c|c|c|c|c|}
\hline
& \multicolumn{5}{|@{}c@{}|}{$\lambda_u^{(d)}$} & \multicolumn{8}{|@{}c@{}|}{$\lambda_t^{(d)}$}  \\
\hline
Amplitude & $T$ & $C$ & $E$ & $P_{uc}$ & $PA_{uc}$ & $P_{tc}$ & $PA_{tc}$ & $P_{EW}$ & $P_{EW}^C$ & $P_{EW}^A$ &  $P_{EW}^E$ & $P_{EW}^{P_u}$ & $P_{EW}^{PA_u}$ \\
\hline
$\sqrt{2} A^{+0}$ & $-1$ & $-1$ & 0 & 0 & 0 & 0 & 0 & $-1$ & $-1$ & $0$ & 0 & 0 & 0 \\
$A^{+-}$ & $-1$ & 0 & $-1$ & $-1$ & $-1$ & $-1$ & $-1$ & 0 & $-\frac{2}{3}$ & $-\frac{1}{3}$ & $\frac{1}{3}$ & $\frac{1}{3}$ & $\frac{1}{3}$ \\
$\sqrt{2} A^{00}$ & 0 & $-1$ & 1 & 1 & 1 & 1 & 1 & $-1$ & $-\frac{1}{3}$ & $\frac{1}{3}$ & $-\frac{1}{3}$ & $-\frac{1}{3}$ & $-\frac{1}{3}$ \\\hline
\end{tabular}
\vspace{0.1in}
\caption{Decomposition of the $B \to \pi\pi$ decay amplitudes in terms of diagrams.}
\label{tab:isopsindiagdeltaS0pipi}
\end{table}

Note that the indirect CP asymmetry in $B^0\to\pi^+\pi^-$ involves the factor $e^{-2i\beta}$ from $B^0$-${\bar B}^0$ mixing. This can be taken into account by multiplying all amplitudes by $e^{i\beta}$. With this in mind, the $B \to \pi\pi$ amplitudes can be written in terms of effective diagrams as
\bea
\sqrt{2} A^{+0} &=& -|V_{ub} V_{ud}| e^{-i\alpha} \, T^{+0} + |V_{tb} V_{td}| \, P_{EW}^{+0} ~, \nn\\
A^{+-} &=& -|V_{ub} V_{ud}| e^{-i\alpha} \, T^{+-} + |V_{tb} V_{td}| \, P_{EW}^{+-} ~, \nn\\
\sqrt{2} A^{00} &=& -|V_{ub} V_{ud}| e^{-i\alpha} \, (T^{+0} - T^{+-}) + |V_{tb} V_{td}| \, (P_{EW}^{+0} - P_{EW}^{+-}) ~.
\eea
The CP-conjugate amplitudes can be obtained from the above by changing the sign of the weak phase $\alpha$.
Here, the effective diagrams $T^{+0}$ and $T^{+-}$ are the combinations of diagrams that are proportional to $\lambda_u^{(d)}$ in $\sqrt{2} A^{+0}$ and $A^{+-}$, respectively. Similarly, $P_{EW}^{+0}$ and $P_{EW}^{+-}$ are the combinations of diagrams that are proportional to $\lambda_t^{(d)}$ in $\sqrt{2} A^{+0}$ and $A^{+-}$, respectively. Consulting Table \ref{tab:isopsindiagdeltaS0pipi}, we see that $T^{+0} = -(T + C)$, while $P_{EW}^{+0} = -(P_{EW} + P_{EW}^C)$. (Note that $T^{+-}$ and $P^{+-}_{EW}$ contain gluonic penguin diagrams.)

The key point here is that $P_{EW}^{+0}$ is related to $T^{+0}$ by the $B \to\pi\pi$ EWP-tree relation of Eq.~(\ref{ETR_isospin}). The amplitudes and CP-conjugate amplitudes are therefore written in terms of three effective diagrams, $T^{+0}$, $T^{+-}$, and $P_{EW}^{+-}$. This implies that there are six parameters: the three magnitudes and two relative strong phases of the effective diagrams, and $\alpha$.

As shown in Sec.~\ref{Bpipi}, there are two RMEs: $A_{1/2,0} \equiv \langle 0 || H_W^{1/2} || \frac12 \rangle$ and $A_{3/2,2} \equiv \langle 2 || H_W^{3/2} || \frac12 \rangle$. In terms of these RMEs, the amplitudes can be written as
\bea
\sqrt{2} A^{+0} &=& \sqrt{\frac32} A_{3/2,2} ~, \nn\\
A^{+-} &=& \frac{1}{\sqrt6} A_{3/2,2} + \frac{1}{\sqrt3} A_{1/2,0} ~, \nn\\
\sqrt{2} A^{00} &=& \sqrt{\frac23}A_{3/2,2} - \frac{1}{\sqrt3}A_{1/2,0} ~.
\eea

Turning to the parameters in the amplitudes, there are a number of issues involved. First, one also has to consider the amplitudes for the CP-conjugate decays. These are usually obtained by simply changing the signs of the weak phases. Second, the indirect CP asymmetry in $B^0\to\pi^+\pi^-$ involves the factor $e^{-2i\beta}$ from $B^0$-${\bar B}^0$ mixing. This can be taken into account by multiplying all amplitudes (CP-conjugate amplitudes) by $e^{i\beta}$ ($e^{-i\beta}$).

Keeping these two points in mind, we note that both RMEs can be written in terms of two pieces, one proportional to $\la_u^{(d)} \equiv V_{ub}^* V_{ud} = |V_{ub} V_{ud}| e^{i\gamma}$, the other proportional to $\la_t^{(d)} \equiv V_{tb}^* V_{td} = |V_{tb} V_{td}| e^{-i\beta}$. The $\la_u^{(d)}$ pieces are generated by the tree operators. For the $A_{1/2,0}$ RME, its $\la_t^{(d)}$ piece gets contributions from both the gluonic penguin and EWP operators. On the other hand, the $\la_t^{(d)}$ piece of the $A_{3/2,2}$ RME is generated only by EWP operators, so there is an EWP-tree relation [Eq.~(\ref{BpipiETR})]. For this reason, we leave $A_{1/2,0}$ as a single term, but write $A_{3/2,2}$ with two pieces. Multiplying the RMEs by $e^{i\beta}$, this gives
\beq
A_{1/2,0} = |A_{1/2,0}| e^{i\phi} ~~,~~~~
A_{3/2,2} = -|V_{ub} V_{ud}| e^{-i\alpha} A^u_{3/2,2} + |V_{tb} V_{td}| A^t_{3/2,2} ~.
\eeq
In the above, $\alpha$ is a weak phase, but the phase $\phi$ is a complicated function of weak and strong phases. $A^t_{3/2,2}$ is related to $A^u_{3/2,2}$ by the EWP-tree relation of Eq.~(\ref{BpipiETR}). The two pieces have the same strong phase, which can be taken to be 0. The CP-conjugate amplitude of $A_{3/2,2}$ is straightforward to obtain -- one simply changes the sign of $\alpha$. However, for the CP conjugate of $A_{1/2,0}$, we have to introduce two new parameters:
\beq
{\bar A}_{1/2,0} = |{\bar A}_{1/2,0}| e^{i{\bar\phi}} ~,
\eeq
where ${\bar\phi}$ is also a (different) complicated function of weak and strong phases. The upshot is that the amplitudes and CP-conjugate amplitudes involve six parameters: $\alpha$, $A^u_{3/2,2}$, $|A_{1/2,0}|$, $|{\bar A}_{1/2,0}|$, $\phi$, and ${\bar\phi}$.

The way this analysis has been carried out up to now is that the EWP diagrams $P_{EW}$ and $P_{EW}^C$  have simply been neglected. The justification is that these contributions are expected to be small (see, for example, GPY \cite{Gronau:1998fn}). If the EWP contribution is neglected, then the direct CP asymmetry in $B^+ \to\pi^+\pi^0$ vanishes since there is only a single contributing amplitude. So there are six observables. But there are also six parameters, which means that the CP phase $\alpha$ can be extracted\footnote{The idea here is that, with six equations in six unknowns, one can solve for the unknown parameters. However, because the equations are highly nonlinear, there will in general be several solutions. Other information, external to the $B \to \pi\pi$ system, will be necessary to resolve this discrete ambiguity.}.

Even so, there is a theoretical uncertainty associated with this determination since the EWP diagrams have been neglected. This uncertainty may be small, but it can be avoided, given that the EWP contributions can be related to the tree contributions. Above, we showed that, even if one includes the EWP $P_{EW}^{+0}$ term, there are still six parameters. And since $A^u_{3/2,2}$ and $A^t_{3/2,2}$ have the same strong phase,
the direct CP asymmetry in $B^+ \to\pi^+\pi^0$ vanishes, so there are still six observables. Thus, once again, $\alpha$ can be extracted (albeit with discrete ambiguities). But the theoretical error has been removed.

We must admit that this observation is not completely new. The $B \to \pi\pi$ SU(2)$_I$ EWP-tree relation of Eq.~(\ref{BpipiETR}) was pointed out by GPY \cite{Gronau:1998fn}. However, because, at that time, $\alpha$ was extracted from $B\to\pi\pi$ decays using a geometrical construction, it was not obvious that $\alpha$ could still be obtained even with the inclusion of the EWP contributions. In 2017, the CKMfitter group performed an update to the extraction of $\alpha$ from $B\to\pi\pi$ and related decays \cite{Charles:2017evz}. Interestingly, they were aware of the $B\to\pi\pi$ EWP-tree relation -- it was mentioned explicitly in the paper -- but instead of integrating it into the analysis, they put an upper limit on the size of the EWP contributions. In the present paper, our point is simply to note that the EWP contributions can be included when extracting $\alpha$ from a fit to the $B\to\pi\pi$ data.

\section{\boldmath The $\btopik$ puzzle}
\label{section:btopik}
Another set of decays whose amplitudes are related by isospin, and which has been analyzed on its own, is $\btopik$. There are four such decays, $B^+ \to \pi^+ K^0$, $B^+ \to \pi^0 K^+$, $\bd\to \pi^- K^+$ and $\bd\to\pi^0 K^0$; they were discussed in Sec.~\ref{BpiKsection}. Their amplitudes satisfy the isospin quadrilateral relation given in Eq.~(\ref{eq:isoquad}).

There are nine observables in these decays: the four branching ratios, the four direct CP asymmetries, and the indirect CP asymmetry in $\bd\to \pi^0K^0$. Following the first observations of $\btopik$ modes by CLEO in 1997 \cite{CLEO:1997ree}, these nine observables have been measured by many experiments, including recent measurements by Belle-II \cite{Belle-II:2023ksq}.

In 2003, it was noted that there is an inconsistency among all the $\btopik$ observables  \cite{Buras:2003yc, Buras:2003dj, Buras:2004ub}. This was dubbed the ``$\btopik$ puzzle.'' In Ref.~\cite{Baek:2004rp} (2004), it was pointed out that one can quantify the discrepancy with the SM by performing a full fit to the $\btopik$ data. Since then, such a fit has been performed many times, sometimes to update the $\btopik$ puzzle \cite{Baek:2007yy, Baek:2009pa, Beaudry:2017gtw}, sometimes to study it in the context of a particular type of new physics \cite{Bhattacharya:2021shk, Datta:2024zrl}. In all cases, the fit was performed using the magnitudes and phases of diagrams as the theoretical unknown parameters, and the SU(3)$_F$ EWP-tree relations were applied.

Above, we have argued that, since the $\btopik$ amplitudes are related by SU(2)$_I$, and not the full SU(3)$_F$, the $\btopik$ SU(2)$_I$ EWP-tree relations given in Eq.~(\ref{eq:ETRBPiK}) are more appropriate for the fit to the $\btopik$ data. In this section, we discuss how these EWP-tree relations can be introduced into the analysis of $\btopik$ decays. We show how the result of Sec.~\ref{Rel_observables_RMEs}, that a relation among $\btopik$ observables is exact under SU(2)$_I$, can be understood in terms of diagrams. And in particular, we redo the fits to the $\btopik$ data and compare the results when the SU(3)$_F$ and $\btopik$ SU(2)$_I$ EWP-tree relations are used.

\subsection{\bf\boldmath Parametrizations of the $\btopik$ amplitudes}
\label{Parametrizations}

We designate the four $\btopik$ amplitudes as $A^{ij}$, where the indices $i$ and $j$ indicate the charges of the final-state $\pi$ and $K$, respectively. The expressions for these amplitudes in terms of diagrams are
\bea
A^{+0} &=& \lambda_{u}^{(s)} \left[P_{uc} + A\right]
+ \lambda_{t}^{(s)} \left[P_{tc} - \frac{1}{3}P^C_{EW} + \frac{2}{3} P^E_{EW} - \frac{1}{3}P^{P_u}_{EW} \right] ~, \nn\\
\s A^{0+} &=& \lambda_{u}^{(s)} \left[-T - C - P_{uc} - A\right]
 +~\lambda_{t}^{(s)} \left[ -P_{tc}-P_{EW}-\frac{2}{3}P_{EW}^C -\frac{2}{3}P_{EW}^E + \frac{1}{3}P^{P_u}_{EW}\right] ~, \nn\\
A^{-+} &=& \lambda_{u}^{(s)} \left[- T - P_{uc}\right]
+ \lambda_{t}^{(s)} \left[ - P_{tc} - \frac{2}{3}P_{EW}^C + \frac{1}{3}P^E_{EW} + \frac{1}{3}P^{P_u}_{EW} \right] ~, \nn\\
\s A^{00} &=& \lambda_{u}^{(s)} \left[-C + P_{uc} \right]  +~\lambda_{t}^{(s)} \left[P_{tc} - P_{EW} - \frac{1}{3}P_{EW}^C - \frac{1}{3}P^E_{EW} - \frac{1}{3}P^{P_u}_{EW} \right] ~.
\label{BpiKamps}
\eea
The amplitudes for the CP-conjugate processes can be obtained from the above by changing the sign of the weak phases.

Even though four (five) diagrams proportional to $\lambda_{u}^{(s)}$ ($\lambda_{t}^{(s)}$) contribute to the $\btopik$ amplitudes, these amplitudes involve only three SU(2)$_I$ RMEs (see Sec.~\ref{BpiKsection}). One can therefore construct three effective diagrams proportional to each of $\lambda_{u}^{(s)}$ and $\lambda_{t}^{(s)}$. This is done by absorbing the diagrams $A$, $P_{EW}^E$, and $P_{EW}^{P_u}$ into the remaining six diagrams through the following redefinitions:
\bea
&\widetilde{T} \equiv T - A  ~,~~\widetilde{C} \equiv C + A  ~,~~ \widetilde{P_{uc}} \equiv P_{uc} + A  ~,~  \nn \\
\widetilde{P_{EW}} \equiv P_{EW} + &P_{EW}^E  ~,~~ \widetilde{P_{EW}^C} \equiv P_{EW}^C - P_{EW}^E  ~,~~  \widetilde{P_{tc}} \equiv P_{tc} - \frac{1}{3} (P_{EW}^{P_u}  - P_{EW}^E)  ~.
\label{BpiKeffdiags}
\eea
In terms of the redefined diagrams, the $\btopik$ amplitudes take the following form:
\bea
A^{+0} &=& \lambda_{u}^{(s)} \widetilde{P_{uc}}
+ \lambda_{t}^{(s)} \left[\widetilde{P_{tc}} - \frac{1}{3} \widetilde{P^C_{EW}} \right] ~, \nn\\
\s A^{0+} &=& \lambda_{u}^{(s)} \left[-\widetilde{T} - \widetilde{C} - \widetilde{P_{uc}} \right]
 +~\lambda_{t}^{(s)} \left[-\widetilde{P_{tc}}-\widetilde{P_{EW}}-\frac{2}{3}\widetilde{P_{EW}^C}\right] ~, \nn\\
A^{-+} &=& \lambda_{u}^{(s)} \left[- \widetilde{T} - \widetilde{P_{uc}}\right]
+ \lambda_{t}^{(s)} \left[ - \widetilde{P_{tc}} - \frac{2}{3}\widetilde{P_{EW}^C} \right] ~, \nn\\
\s A^{00} &=& \lambda_{u}^{(s)} \left[-\widetilde{C} + \widetilde{P_{uc}} \right]  +~\lambda_{t}^{(s)} \left[\widetilde{P_{tc}} - \widetilde{P_{EW}} - \frac{1}{3} \widetilde{P_{EW}^C} \right] ~.
\label{BpiKampsred}
\eea

These expressions for the $\btopik$ amplitudes in terms of diagrams hold regardless of whether the underlying symmetry is assumed to be SU(2)$_I$ or SU(3)$_F$. However, the EWP-tree relations in the two cases are different. Let us examine how they can be connected to the above diagrammatic parametrization.

As pointed out in Sec.~\ref{BpiKsection}, there are two $\Delta S = 1$ SU(2)$_I$ EWP-tree relations involving the $\btopik$ diagrams, given in Eq.~(\ref{eq:ETRBPiK}). We note that these EWP-tree relations can be rewritten using the redefinitions given in Eq.~(\ref{BpiKeffdiags}):
\beq
\widetilde{P_{EW}} = -\frac{3}{2}\frac{c_9+c_{10}}{c_1+c_2}\,\widetilde{C} ~~,~~~~
\widetilde{P_{EW}^{C}} = -\frac{3}{2}\frac{c_9+c_{10}}{c_1+c_2}\,\widetilde{T}~.
\label{SU2etrBpik}
\eeq

Thus, under SU(2)$_I$ symmetry, the four $\btopik$ decays can be parametrized using six diagrams, of which only four are independent when taking the EWP-tree relations into account. This means that seven real hadronic parameters describe the system: four magnitudes of diagrams and three relative strong phases. With nine observables, a fit can be done.

Turning to SU(3)$_F$, the relevant EWP-tree relations are given in the first two lines of Eq.~(\ref{eq:su3_ETR_relations}), repeated here for convenience:
\bea
P_{EW} + P_{EW}^E &=& -\frac{3}{4}\left[\frac{(c_9 + c_{10})}{(c_1 + c_2)}\left(T+C+A+E\right)+\frac{(c_9 - c_{10})}{(c_1 - c_2)} \left(T-C-A+E\right)\right]\,, \nn\\
P_{EW}^C - P_{EW}^E &=& -\frac{3}{4}\left[\frac{(c_9 + c_{10})}{(c_1 + c_2)}\left(T+C-A-E\right)-\frac{(c_9 - c_{10})}{(c_1 - c_2)} \left(T-C-A+E\right)\right]\,.~~~~~
\label{SU3etrBpik}
\eea
If one attempts to use the diagram redefinitions of Eq.~(\ref{BpiKeffdiags}) to rewrite these EWP-tree relations, one sees immediately that there are problems. These relations include the exchange diagram $E$, which does not appear in the $\btopik$ amplitudes [Eq.~(\ref{BpiKamps})]. Furthermore, one finds that, although the redefinitions applied to the left-hand sides of these equations remove $P^E_{EW}$, the annihilation diagram $A$ cannot be readily removed. So it is not obvious how to apply the SU(3)$_F$ EWP-tree relations to the $\btopik$ system.

What is going on here is the following. The consequences of having an underlying SU(3)$_F$ symmetry are very different from those of an SU(2)$_I$ symmetry. With SU(3)$_F$, one must take all charmless $B \to PP$ decays into account; with SU(2)$_I$, only the $\btopik$ decays are considered. Thus, while the $\btopik$ amplitudes are written in terms of three SU(2)$_I$ RMEs, they are written in terms of four SU(3)$_F$ RMEs. Also, SU(3)$_F$ and SU(2)$_I$ diagrams are not the same, since they apply to different sets of decays.
Under SU(3)$_F$, there are once again more diagrams than RMEs that describe the $B\to PP$ decays, so that some diagrams must be removed by making redefinitions similar to those of Eq.~(\ref{BpiKeffdiags}). It is these redefined diagrams that appear in the EWP-tree relations above. However, they are defined considering all charmless $B \to PP$ decays; they generally do not apply if only a subset of decays, such as $\btopik$, is used. It is all of these factors that lead to the appearance of the $A$ and $E$ diagrams in the SU(3)$_F$ EWP-tree relations, and which make it difficult to apply them to $\btopik$ decays.

One way to remove $A$ and $E$ entirely is to add the above two SU(3)$_F$ EWP-tree relations together and obtain the combined relation of Eq.~(\ref{ETR_orig}). In this case, there will be six diagrams and one EWP-tree relation, resulting in nine independent real parameters: five magnitudes of diagrams and four relative strong phases. With nine observables, in principle, a fit can be performed.
The problem with such a fit is that there are zero degrees of freedom. If the equations for the observables as a function of the unknown parameters were linear, there would be a unique solution. However, the equations are nonlinear, so that, in general, it is likely that there will be multiple discretely-ambiguous solutions. In addition, without any theory input, the size of the $E$ diagram (which doesn't appear in the $\btopik$ amplitudes, but does appear in the EWP-tree relations) is allowed to be large. For these reasons, this is not an ideal scenario.

An alternative (and more useful) route is to add theory input. In Ref.~\cite{Gronau:1994rj}, it is argued that, since the diagrams $E$, $A$, and $PA$ all require the initial ${\bar b}$ quark to interact with the spectator quark, they are expected to be quite suppressed compared to $T$ and $C$. This is borne out experimentally: the amplitudes for $B_s^0\to\pi^+\pi^-$ and $B^+ \to D^+ K^0$ are dominated by $E + PA$ and $A$, respectively; and their branching ratios are measured to be considerably smaller than those of other $B$ decays whose amplitudes are dominated by $T$ or $C$. It is therefore a reasonable approximation to neglect $E$, $A$, and $PA$ (as well as their EWPs).

Neglecting $E$, $A$, and $P_{EW}^E$ leads to the following modified SU(3)$_F$ EWP-tree relations:
\bea
P_{EW} &=& -\frac{3}{4}\left[\frac{(c_9 + c_{10})}{(c_1 + c_2)}\left(T+C\right)+\frac{(c_9 - c_{10})}{(c_1 - c_2)} \left(T-C\right)\right]\,, \label{SU3etrBpikm1} \nn\\
P_{EW}^C &=& -\frac{3}{4}\left[\frac{(c_9 + c_{10})}{(c_1 + c_2)}\left(T+C\right)-\frac{(c_9 - c_{10})}{(c_1 - c_2)} \left(T-C\right)\right]\,.
\label{SU3etrBpikm2}
\eea
Under SU(3)$_F$ ($+$ theory input), we now have a situation very much like that of SU(2)$_I$: there are six diagrams and two EWP-tree relations. This corresponds to seven real parameters, so a fit can be done. Indeed, all of the previous $\btopik$ fits \cite{Baek:2004rp, Baek:2007yy, Baek:2009pa, Beaudry:2017gtw, Bhattacharya:2021shk, Datta:2024zrl} have applied the above EWP-tree relations to the $\btopik$ system. In Sec.~(\ref{sec:bpikfits}) below, we compare the results of the fits using the $\btopik$ SU(2)$_I$ and SU(3)$_F$ EWP-tree relations, respectively Eqs.~(\ref{SU2etrBpik}) and (\ref{SU3etrBpikm2}).

We note in passing that it is not really necessary to neglect all of $E$, $A$, and $PA$. Beginning with the full SU(3)$_F$ EWP-tree relations of Eq.~(\ref{SU3etrBpik}), if we make the redefinitions of Eq.~(\ref{BpiKeffdiags}) and add the redefinition\footnote{In Ref.~\cite{Bhattacharya:2025wcq}, SU(3)$_F$ symmetry was used to study all $B\to PP$ decays. There, the decay amplitudes were written in terms of effective diagrams by removing $E$. However, $E$ does not appear in the $\btopik$ amplitudes, so a natural choice for redefining diagrams here is to eliminate $A$ instead.} $\widetilde{E} \equiv E + A$, the SU(3)$_F$ EWP-tree relations can be recast as
\bea
\widetilde{P_{EW}} &=& -\frac{3}{4}\left[\frac{(c_9 + c_{10})}{(c_1 + c_2)}\left(\widetilde{T}+\widetilde{C}+\widetilde{E}\right)+\frac{(c_9 - c_{10})}{(c_1 - c_2)} \left(\widetilde{T}-\widetilde{C}+\widetilde{E}\right)\right]\,, \label{SU3etrBpikum1} \nn\\
\widetilde{P_{EW}^C} &=& -\frac{3}{4}\left[\frac{(c_9 + c_{10})}{(c_1 + c_2)}\left(\widetilde{T}+\widetilde{C}-\widetilde{E}\right)-\frac{(c_9 - c_{10})}{(c_1 - c_2)} \left(\widetilde{T}-\widetilde{C}+\widetilde{E}\right)\right]\,.
\label{SU3etrBpikum2}
\eea
In order to recover the modified SU(3)$_F$ EWP-tree relations of Eq.~(\ref{SU3etrBpikm2}), we see that it is only necessary to neglect the redefined $\widetilde{E}$ diagram. (This said, it is difficult to come up with a theoretical justification for neglecting only this diagram, and not all of $E$, $A$, and $PA$.)

\subsection{\bf\boldmath A relation among $\btopik$ observables (diagrams)}

In Sec.~\ref{Rel_observables_RMEs}, we defined the following quantity involving observables in $\btopik$ decays:
\bea
\de_{\pi K} &=& \frac{{\cal B}(\pi^-K^+)A_{CP}(\pi^-K^+)}{\Pi(\pi^-K^+)} + \frac{{\cal B}(\pi^+K^0)A_{CP}(\pi^+K^0)}{\Pi(\pi^+K^0)} \nn \\
&&\hspace{2truecm} -~\frac{2\,{\cal B}(\pi^0K^+)A_{CP}(\pi^0K^+)}{\Pi(\pi^0K^+)} - \frac{2\,{\cal B}(\pi^0K^0)A_{CP}(\pi^0K^0)}{\Pi(\pi^0K^0)}\,.
\eea
Here, ${\cal B}(\pi^i K^j)$, $A_{CP}(\pi^i K^j)$ and $\Pi(\pi^i K^j)$ are respectively the CP-averaged branching ratio, the direct CP asymmetry and the two-body phase-space factor for the decay $B \to\pi^iK^j$. In Ref.~\cite{Gronau:2005kz}, it was argued that, if one assumes SU(3)$_F$ symmetry and one adds some theory input regarding the magnitudes and phases of certain diagrams, one finds $\de_{\pi K} \approx 0$.
In Sec.~\ref{Rel_observables_RMEs}, we showed that $\de_{\pi K} = 0$ is an exact consequence of the $\btopik$ SU(2)$_I$ EWP-tree relations; no theory input is necessary. This was demonstrated using RMEs, but it is useful to revisit this proof, this time in terms of diagrams.

Using the redefined diagrams of Eq.~(\ref{BpiKeffdiags}), we find
\bea
\de_{\pi K} &=& -4\,{\rm Im}\[\la^{(s)}_u\la^{(s)*}_t\]{\rm Im}\[\(\widetilde{P_{EW}}+\widetilde{P_{EW}^C}\)^*\(\widetilde{T}+\widetilde{C}\) + \(\widetilde{P_{EW}^*} \widetilde{C} - \widetilde{P_{EW}^{C*}} \widetilde{T}\)\]\,.
\eea
Let us focus on the second term in square brackets, the one dependent on diagrams. For the first piece, $\widetilde{P_{EW}} + \widetilde{P^C_{EW}} \propto \widetilde{T} + \widetilde{C}$ under both SU(3)$_F$ and SU(2)$_I$, see Eqs.~(\ref{SU2etrBpik}) and (\ref{SU3etrBpikum2}). This term is therefore real, i.e., its imaginary component vanishes.

It is the second piece that is particularly interesting. Consider first SU(3)$_F$. The complete EWP-tree relations are given in Eq.~(\ref{SU3etrBpik}). If one takes $(c_9 + c_{10})/(c_1 + c_2) = (c_9 - c_{10})/(c_1 - c_2)$, which holds to about 3\%, and one neglects $\widetilde{E}$, one finds from Eq.~(\ref{SU3etrBpikum2}), $\widetilde{P_{EW}} \propto \widetilde{T}$ and $\widetilde{P_{EW}^C} \propto \widetilde{C}$, with the same proportionality constants. In this limit, the second term is proportional to ${\rm Im}(\widetilde{T^*} \widetilde{C})$. Assuming that $\widetilde{C}$ and $\widetilde{T}$ have the same strong phase \cite{Gronau:2005kz}, this term vanishes. Given all the assumptions and approximations, the most that can be claimed is that we expect $\de_{\pi K} \approx 0$.

Now consider SU(2)$_I$. Its $\btopik$ EWP-tree relations [Eq.~(\ref{SU2etrBpik})] state that $\widetilde{P_{EW}} \propto \widetilde{C}$ and $\widetilde{P_{EW}^C} \propto \widetilde{T}$, also with the same proportionality constants. Thus, $(\widetilde{P_{EW}^*} \widetilde{C} - \widetilde{P_{EW}^{C*}} \widetilde{T}) \propto (|\widetilde{C}|^2 - |\widetilde{T}|^2)$, which is purely real, so that $\de_{\pi K}$ identically vanishes. This demonstrates that $\de_{\pi K} = 0$ is an exact consequence of applying the $\btopik$ SU(2)$_I$ EWP-tree relations to $\btopik$ decays.

\subsection{\bf\boldmath Fits to the $\btopik$ data}
\label{sec:bpikfits}

The $\btopik$ amplitudes can be expressed in terms of six diagrams: $\widetilde{T}$, $\widetilde{C}$, $\widetilde{P_{uc}}$ (proportional to $\lambda_{u}^{(s)}$) and $\widetilde{P_{tc}}$, $\widetilde{P_{EW}}$, $\widetilde{P_{EW}^C}$ (proportional to $\lambda_{t}^{(s)}$). This holds regardless of whether the underlying symmetry group is taken to be SU(2)$_I$ or SU(3)$_F$. The EWP diagrams $\widetilde{P_{EW}}$ and $\widetilde{P_{EW}^C}$ are related to the tree diagrams $\widetilde{T}$ and $\widetilde{C}$ by EWP-tree relations, but these relations are different for the two symmetry groups.
[The $\btopik$ SU(2)$_I$ and SU(3)$_F$ EWP-tree relations used in the fits are given by Eqs.~(\ref{SU2etrBpik}) and (\ref{SU3etrBpikm2}), respectively.]
Once these relations are applied, there are
four independent diagrams in the $\btopik$ amplitudes. This corresponds to seven unknown parameters: four magnitudes of diagrams and three relative strong phases.

As mentioned previously, a total of nine observables have been measured in $\btopik$ decays: the four branching ratios ${\cal B}$, the four direct CP asymmetries $A_{CP}$, and the (mixing-induced) indirect CP asymmetry $S_{CP}$ in $\bd\to \pi^0K^0$. With more observables than parameters, a fit can be done.

The latest $\btopik$ measurements are shown in Table \ref{tab:BpiK_data}. In performing the fits, the weak phases $\gamma$, $\beta$ and $\beta_s$ appear in the theoretical expressions for the observables. The values of these weak phases are fixed to their world averages \cite{ParticleDataGroup:2024cfk}.

\begin{table}[!htbp]
\centering
\begin{tabular}{|c|c|c|c|}
\hline
Mode & ${\cal B}~~[10^{-6}]$ & $A_{CP}$ & $S_{CP}$ \\
\hline
$B^{+} \to \pi^{+} K^{0}$ & $23.9 \pm 0.6$ & $-0.003 \pm 0.015$ & \\
$B^{+} \to \pi^{0} K^{+}$ & $13.2 \pm 0.4$ & $0.027 \pm 0.012$ & \\
$B^{0} \to \pi^{-} K^{+}$ & $20.0 \pm 0.4$ & $-0.0831 \pm 0.0031$ & \\
$B^{0} \to \pi^{0} K^{0}$ & $10.1 \pm 0.4$ & $0.00 \pm 0.08$ & $0.64 \pm 0.13$ \\ \hline
\end{tabular}
\caption{Branching ratios ${\cal B}$, direct CP asymmetries $A_{CP}$, and indirect CP asymmetry $S_{CP}$
 for the four $\btopik$ decay modes. The data are taken from the Particle Data Group \cite{ParticleDataGroup:2024cfk}.}
\label{tab:BpiK_data}
\end{table}

All the fits are done using the {\it Minuit} package \cite{James:1975dr} to find the minimum $\chi^2$ \footnote{The fits have been performed
using the \emph{migrad} method, which returns a symmetric error based on the Hessian evaluated
at the minimum of the $\chi^2$ function.}. In the first fit, we allow the magnitudes of the diagrams to vary freely. (The strong phase of the $\widetilde{T}$ diagram is fixed to 0; the strong phases of other diagrams are allowed to vary between 0 and $360^\circ$.) The results are shown in Table \ref{tab:noconstrained}. The left (right) side of the table shows the results for the case where the SU(3)$_F$ ($\btopik$ SU(2)$_I$) EWP-tree relations are used (the fit with the SU(3)$_F$ relations is an update to Refs.~\cite{Baek:2004rp, Baek:2007yy, Baek:2009pa, Beaudry:2017gtw, Bhattacharya:2021shk, Datta:2024zrl}). Although the SU(3)$_F$ EWP-tree relations provide a somewhat better fit to the data, both fits appear to be perfectly acceptable.

\begin{table} [h!!!!]
\centering
\begin{tabular}{|c|c|c|}
\hline
& \multicolumn{1}{|c|}{SU(3)$_F$ EWP-tree relations} & \multicolumn{1}{|c|}{$\btopik$ SU(2)$_I$ EWP-tree relations} \\ &\multicolumn{1}{|c|}{$\chi^{2}_{\rm min}/\text{d.o.f.} = 0.98/2 $} & \multicolumn{1}{|c|}{$\chi^{2}_{\rm min}/\text{d.o.f.} = 2.9/2$}   \\
& \multicolumn{1}{|c|}{$p$-value = 0.61} & \multicolumn{1}{|c|}{$p$-value = 0.24} \\
\hline
Parameter  & Best-fit value & Best-fit value  \\
\hline
$|\widetilde{T}|$ & $59\pm 5$ & $ 33\pm 13$ \\
\hline
$|\widetilde{C}|$ & $62\pm5$ & $ 40\pm 10$ \\
\hline
$|\widetilde{P_{uc}}|$ & $65.1\pm1.0$ & $ 38\pm 5$ \\
\hline
$|\widetilde{P_{tc}}|$ & $0.66\pm 0.08$ & $ 0.80\pm 0.04$ \\
\hline
$\delta_{\widetilde{C}}$ & $(182.7\pm1.5)^{\circ}$  &  $(180.5\pm1.2)^{\circ}$ \\
\hline
$\delta_{\widetilde{P_{uc}}}$ & $(182.43\pm 0.19)^{\circ}$   & $(175.6 \pm1.6)^{\circ}$  \\
\hline
$\delta_{\widetilde{P_{tc}}}$ & $(182.4\pm0.6 )^{\circ}$   & $(356.3\pm2.0)^{\circ}$  \\
\hline
\end{tabular}
\caption{Fit: magnitudes of diagrams allowed to vary freely, phases vary between 0 and $360^{\circ}$. On the left (right): best-fit values of the unknown parameters for the case where the SU(3)$_F$ [$\btopik$ SU(2)$_I$] EWP-tree relations have been used.}
\label{tab:noconstrained}
\end{table}

However, both fits have serious problems. One of the advantages of diagrams over RMEs is that, while they are not Feynman diagrams, they do represent dynamical ways in which the decay in question can be generated. As a result, one can estimate the relative sizes of various diagrams. (We already saw this in Sec.~\ref{Parametrizations}, where we argued that, from a theoretical point of view, the diagrams $E$, $A$ and $PA$ are expected to be much smaller than other diagrams.) Now consider the ratio $\widetilde{C}/\widetilde{T} = (C + A)/(T - A)$. It is argued that $|A| \ll |T|,|C|$ \cite{Gronau:1994rj}, in which case this ratio is essentially $C/T$.
Naively, it is expected that $|C/T| = 1/3$, simply by counting colors. This is supported by theoretical calculations. This ratio has been computed for $\btopik$ decays within QCD factorization, and $|C/T| \simeq 0.2$ is found at NLO \cite{Beneke:2001ev}. The NNLO analysis gives $0.13 \le |C/T| \le 0.43$, with a central value of $|C/T| = 0.23$, very near its NLO value \cite{Bell:2007tv, Bell:2009nk, Beneke:2009ek, Bell:2015koa}. On the other hand, in the above fits, the best-fit value of $|\widetilde{C}/\widetilde{T}|$ is $1.05 \pm 0.12$ [SU(3)$_F$ EWP-tree relations] or $1.2 \pm 0.6$ [$\btopik$ SU(2)$_I$ EWP-tree relations]. These central values are much larger than the theoretical expectation.

With this in mind, we redo the fits, fixing the ratio $|\widetilde{C}/\widetilde{T}| = 0.2$. The results are shown in Table \ref{tab:C/T0.2}. We now find that the fits are (very) poor. If the SU(3)$_F$ EWP-tree relations are used, the best fit has $\chi^2_{\rm min}/{\rm d.o.f.} = 15.6/3$, for a $p$-value of $1.3\times 10^{-3}$. This corresponds to a 3.2$\sigma$ discrepancy with the SM. And with the $\btopik$ SU(2)$_I$ EWP-tree relations, the fit is even worse: here we find $\chi^2_{\rm min}/{\rm d.o.f.} = 29.2/3$, for a $p$-value of $2.0\times10^{-6}$, which represents a 4.8$\sigma$ discrepancy with the SM.

\begin{table} [h!!!!]
\centering
\begin{tabular}{|c|c|c|}
\hline
& \multicolumn{1}{|c|}{SU(3)$_F$ EWP-tree relations} & \multicolumn{1}{|c|}{$\btopik$ SU(2)$_I$ EWP-tree relations} \\ &\multicolumn{1}{|c|}{$\chi^{2}_{\rm min}/\text{d.o.f.} = 15.6/3$} & \multicolumn{1}{|c|}{$\chi^{2}_{\rm min}/\text{d.o.f.} = 29.2/3$}   \\
& \multicolumn{1}{|c|}{$p$-value $=1.3 \times 10^{-3}$} & \multicolumn{1}{|c|}{$p$-value $=2.0 \times 10^{-6}$} \\
\hline
Parameter  & Best-fit value & Best-fit value  \\
\hline
$|\widetilde{T}|$ & $10.0 \pm 1.1$ & $ 34\pm 5$ \\
\hline
$|\widetilde{P_{uc}}|$ & $2.2\pm 2.9$ & $ 2.1\pm 2.5$ \\
\hline
$|\widetilde{P_{tc}}|$ & $1.232 \pm 0.027$ & $ 1.037\pm 0.023$ \\
\hline
$\delta_{\widetilde{C}}$ & $(256.1\pm17)^{\circ}$  & $( 36\pm6)^{\circ}$  \\
\hline
$\delta_{\widetilde{P_{uc}}}$ & $(395\pm18)^{\circ}$  & $( 183\pm12 )^{\circ}$  \\
\hline
$\delta_{\widetilde{P_{tc}}}$ & $(346.1\pm3.4)^{\circ}$  & $( 185.2\pm0.9)^{\circ}$  \\
\hline
\end{tabular}
\caption{Fit: we have fixed $|\widetilde{C}/\widetilde{T}| = 0.2$, phases vary between 0 and $360^{\circ}$. On the left (right): best-fit values of the unknown parameters for the case where the SU(3)$_F$ [$\btopik$ SU(2)$_I$] EWP-tree relations have been used.}
\label{tab:C/T0.2}
\end{table}

Finally, some might say that the constraint $|\widetilde{C}/\widetilde{T}| = 0.2$ is too stringent. We therefore redo the fits, this time taking $|\widetilde{C}/\widetilde{T}| =0.5$. The results are shown in Table \ref{tab:C/T0.5}. Now the fit that uses the SU(3)$_F$ EWP-tree relations is not bad. However, if the $\btopik$ SU(2)$_I$ EWP-tree relations are used, the fit is still very poor: we find $\chi^2_{\rm min}/{\rm d.o.f.} = 26.2/3$, for a $p$-value of $8.7 \times 10^{-6}$, or a 4.4$\sigma$ discrepancy with the SM.

\begin{table} [h!!!!]
\centering
\begin{tabular}{|c|c|c|}
\hline
& \multicolumn{1}{|c|}{SU(3)$_F$ EWP-tree relations} & \multicolumn{1}{|c|}{$\btopik$ SU(2)$_I$ EWP-tree relations} \\ &\multicolumn{1}{|c|}{$\chi^{2}_{\rm min}/\text{d.o.f.} = 5.6/3 $} & \multicolumn{1}{|c|}{$\chi^{2}_{\rm min}/\text{d.o.f.} = 26.2/3$}   \\
& \multicolumn{1}{|c|}{$p$-value = 0.13} & \multicolumn{1}{|c|}{$p$-value $=8.7 \times 10^{-6}$} \\
\hline
Parameter  & Best-fit value & Best-fit value  \\
\hline
$|\widetilde{T}|$ & $6.9 \pm 0.8$ & $ 13\pm 4$ \\
\hline
$|\widetilde{P_{uc}}|$ & $1.8 \pm2.5$ & $ 9\pm 4$ \\
\hline
$|\widetilde{P_{tc}}|$ & $1.23\pm 0.04$ & $ 1.229\pm 0.035$ \\
\hline
$\delta_{\widetilde{C}}$ & $( 224\pm15)^{\circ}$  & $( 162\pm7)^{\circ}$  \\
\hline
$\delta_{\widetilde{P_{uc}}}$ & $( 340\pm19)^{\circ}$  & $( 190.8\pm3.4)^{\circ}$  \\
\hline
$\delta_{\widetilde{P_{tc}}}$ & $( 340\pm 5 )^{\circ}$  & $( 192\pm4)^{\circ}$  \\
\hline
\end{tabular}
\caption{Fit: we have fixed $|\widetilde{C}/\widetilde{T}| = 0.5$, phases vary between 0 and $360^{\circ}$. On the left (right): best-fit values of the unknown parameters for the case where the SU(3)$_F$ [$\btopik$ SU(2)$_I$] EWP-tree relations have been used.}
\label{tab:C/T0.5}
\end{table}

We therefore conclude the following. The $\btopik$ puzzle has been around for some twenty years. Up to now, fits to the $\btopik$ data have always used the SU(3)$_F$ EWP-tree relations ($+$ theory input), and discrepancies with the SM at the level of 2-3$\sigma$ were found. However, since the $\btopik$ amplitudes are related to one another by isospin, and not the full SU(3)$_F$, the $\btopik$ SU(2)$_I$ EWP-tree relations should really be used. When one does this (and adds some theory input), it is found that the $\btopik$ puzzle actually presents a 4-5$\sigma$ discrepancy with the SM, much larger than was previously realized.

\section{Conclusions}
\label{section:conclusion}

If one assumes flavor SU(3) symmetry, the amplitudes of all charmless $B \to PP$ decays are related, in that they can all be written as functions of the same SU(3)$_F$ reduced matrix elements. In 1998, it was shown that some RMEs involving electroweak penguin operators are related to those involving tree operators, and that EWP diagrams are related to tree diagrams \cite{Gronau:1998fn}. Recently, Refs.~\cite{Berthiaume:2023kmp, Bhattacharya:2025wcq} performed fits to all the $B \to PP$ data under the assumption of SU(3)$_F$. These SU(3)$_F$ EWP-tree relations were used in the analysis, and a  significant discrepancy with the SM was found.

Consider $\btopik$ decays. The amplitudes of the four processes $B^+ \to \pi^+ K^0$, $B^+ \to \pi^0 K^+$, $\bd\to \pi^- K^+$ and $\bd\to\pi^0 K^0$ obey an isospin quadrilateral relation. After Ref.~\cite{Gronau:1998fn} appeared, the SU(3)$_F$ EWP-tree relations were mostly used in the analysis of $\btopik$ decays.  First, it was shown in Refs.~\cite{Buras:2003yc, Buras:2003dj, Buras:2004ub} that the observables in these decays are not completely consistent with one another. This ``$\btopik$ puzzle'' can be quantified by performing a fit to all the $\btopik$ data. To do this, the $\btopik$ amplitudes are written in terms of diagrams, the SU(3)$_F$ EWP-tree relations are imposed, and the magnitudes and phases of the diagrams are taken to be the free parameters in the fits. The fits performed over the years have typically found that the data disagree with the SM at the level of 2-3$\sigma$ \cite{Baek:2004rp, Baek:2007yy, Baek:2009pa, Beaudry:2017gtw, Bhattacharya:2021shk, Datta:2024zrl}. Second, using the SU(3)$_F$ EWP tree-relations ($+$ theory input), it was shown in Ref.~\cite{Gronau:2005kz} that a relation among $\btopik$ observables holds approximately. This was touted as a test of the SM, and measurements have found that the data reproduce this relationship.

But there is a bit of an inconsistency here: the $\btopik$ amplitudes are related only by isospin, but the EWP-tree relations assume the full SU(3)$_F$ symmetry. In this paper, we show that, if only SU(2)$_I$ is assumed, there are still EWP-tree relations. There are six sets of $B \to PP$ decays whose amplitudes are related by isospin, three with $\Delta S=0$ and three with $\Delta S=1$. In general, each set of decays has its own  SU(2)$_I$ EWP-tree relations. The SU(2)$_I$ EWP-tree relations for $\Delta S = 0$ decays are similar to those of SU(3)$_F$. But the SU(2)$_I$ EWP-tree relations for $\Delta S = 1$ decays are quite different from those of SU(3)$_F$.

This can be understood by looking at the quark-level transitions ${\bar b} \to {\bar d} u {\bar u}$ ($\Delta S=0$) and ${\bar b} \to {\bar s} u {\bar u}$ ($\Delta S=1$). The SU(3)$_F$ weak Hamiltonian for both decays transforms as the product of three fundamental representations, as does the SU(2)$_I$ weak Hamiltonian for $\Delta S=0$ decays. But the SU(2)$_I$ weak Hamiltonian for $\Delta S=1$ decays transforms differently, as the product of two fundamental representations and a singlet. This leads to different EWP-tree relations.

If one analyzes a set of hadronic $B$ decays whose amplitudes are related by isospin, we argue that the SU(2)$_I$ EWP-tree relations for that set of decays must be used in the analysis. The $B \to\pi\pi$ SU(2)$_I$ EWP-tree relations can be used to include EWP contributions in the extraction of $\alpha$ from $B \to \pi\pi$ decays. Using the SU(2)$_I$ EWP-tree relations for $\btopik$, we show that the relation among $\btopik$ observables that was found to hold approximately using the SU(3)$_F$ EWP-tree relations ($+$ theory input) is actually exact. And if these EWP-tree relations are used to analyze the $\btopik$ puzzle, the discrepancy with the SM is found to be at the level of 4-5$\sigma$, much larger than what was found previously.

\acknowledgments{We thank R. Fleischer and G. Paz for respectively bringing Refs.~\cite{Fleischer:2008wb, Fleischer:2018bld} and Ref.~\cite{Paz:2002ev} to our attention.
D.L. thanks S. Kumbhakar for helpful communications. This work was financially supported by the National Science Foundation, Grant No.\ PHY-2310627 (B.B.), by NSERC of Canada (M.B., A.J., D.L., I.R.), and by FRQNT, Scholarship No.\ 363240 (M.B.).}

\bibliography{JHEPrefs}
\bibliographystyle{JHEP}

\end{document}